\documentclass[%
reprint,
nofootinbib,
 amsmath,amssymb,
 prl
floatfix,
]{revtex4-1}

\usepackage{graphicx}
\usepackage{dcolumn}
\usepackage{bm}
\usepackage{graphicx}
\usepackage{enumitem, amsmath,bm}
\usepackage{bbold}
\usepackage{comment}
\usepackage[dvipsnames]{xcolor}
\usepackage{mathtools}
\usepackage{float}
\usepackage{subfig}

\usepackage{scalerel,stackengine}
\stackMath
\newcommand\reallywidehat[1]{%
\savestack{\tmpbox}{\stretchto{%
  \scaleto{%
    \scalerel*[\widthof{\ensuremath{#1}}]{\kern.1pt\mathchar"0362\kern.1pt}%
    {\rule{0ex}{\textheight}}
  }{\textheight}%
}{2.4ex}}%
\stackon[-6.9pt]{#1}{\tmpbox}%
}
\newcommand\utot{\boldsymbol{u}}
\newcommand\ubar{\overline{\boldsymbol{u}}}
\newcommand\uprime{\boldsymbol{u}^\prime}
\newcommand\uhat{\boldsymbol{\whu}}

\newcommand\ujhat{\whu_j}

\newcommand{\be}{\mathbf{e}}
\newcommand{\dt}{\Delta t}
\newcommand{\whu}{\widehat{u}}
\newcommand{\whR}{\widehat{R}}
\newcommand{\whp}{\widehat{p}}
\newcommand{\pd}{\partial}
\newcommand{\bk}{\mathbf{k}}

\newcommand{\strm}{{\rm{strm}}}
\newcommand{\spwise}{{\rm{span}}}
\newcommand{\xt}{x}
\newcommand{\zt}{z}

\begin{document}

\preprint{APS/123-QED}

\title{Phase Transition to Turbulence via Moving Fronts
}
\author{S\'ebastien Gom\'e$^{{1}, {2}}$}
\email{sebastien.gome@gmail.com}
\author{Ali\'enor Rivi\`ere$^{1}$}
\author{Laurette S. Tuckerman$^{1}$}
\author{Dwight Barkley$^{3}$}

\affiliation{
$^{1}$Laboratoire de Physique et M\'ecanique des Milieux H\'et\'erog\`enes, CNRS, ESPCI Paris, PSL Research
University, Sorbonne Universit\'e, Universit\'e Paris-Cit\'e, Paris 75005, France\\
$^{2}$Department of Physics, Technion Israel Institute of Technology, 32000 Haifa, Israel\\
$^{3}$Mathematics Institute, University of Warwick, Coventry CV4 7AL, United Kingdom
}

\date{\today}

\begin{abstract}

Directed percolation (DP), a universality class of continuous phase transitions, has recently been established as a possible route to turbulence in subcritical wall-bounded flows. 
In canonical straight pipe or planar flows, the transition occurs via discrete large-scale turbulent structures, known as puffs in pipe flow or bands in planar flows, which either self-replicate or laminarize. 
However, these processes might not be universal to all subcritical shear flows.
Here, we design a numerical experiment that eliminates 
discrete structures in plane Couette flow and show that it follows a different, simpler transition scenario: turbulence proliferates via expanding fronts 
and decays via spontaneous creation of laminar zones. 
We map this phase transition onto a stochastic one-variable system. 
The level of turbulent fluctuations dictates whether moving-front transition is discontinuous, or continuous and within the DP universality class, with profound implications for other hydrodynamic systems.

\end{abstract}

\maketitle

Landau and Liftshitz \citep{landau1987fluid} describe the subcritical transition to turbulence in shear flows such as pipe and channels as a first-order transition, driven by the competition between two states, one laminar and one turbulent. 
The stable state invades the metastable state via moving fronts. Below a critical 
Reynolds number $Re_c$, laminar flow is the stable asymptotic state. Above $Re_c$, turbulence is stable and will expand into laminar flow. 
\citet{pomeau} recognized that the problem of subcritical transition is richer than this, since turbulence is a {\em fluctuating state} and can spontaneously decay to the laminar state, forming {\em laminar gaps} within turbulence, while laminar flow is an {\em absorbing state} that cannot spontaneously become turbulent. 
The subcritical transition to turbulence is thus
an absorbing state transition that could be second-order and belong to a universality class of non-equilibrium statistical systems known as directed percolation (DP) \citep{grassberger1981phase, janssen1981nonequilibrium}.

\begin{figure}
\includegraphics[width=1\columnwidth]{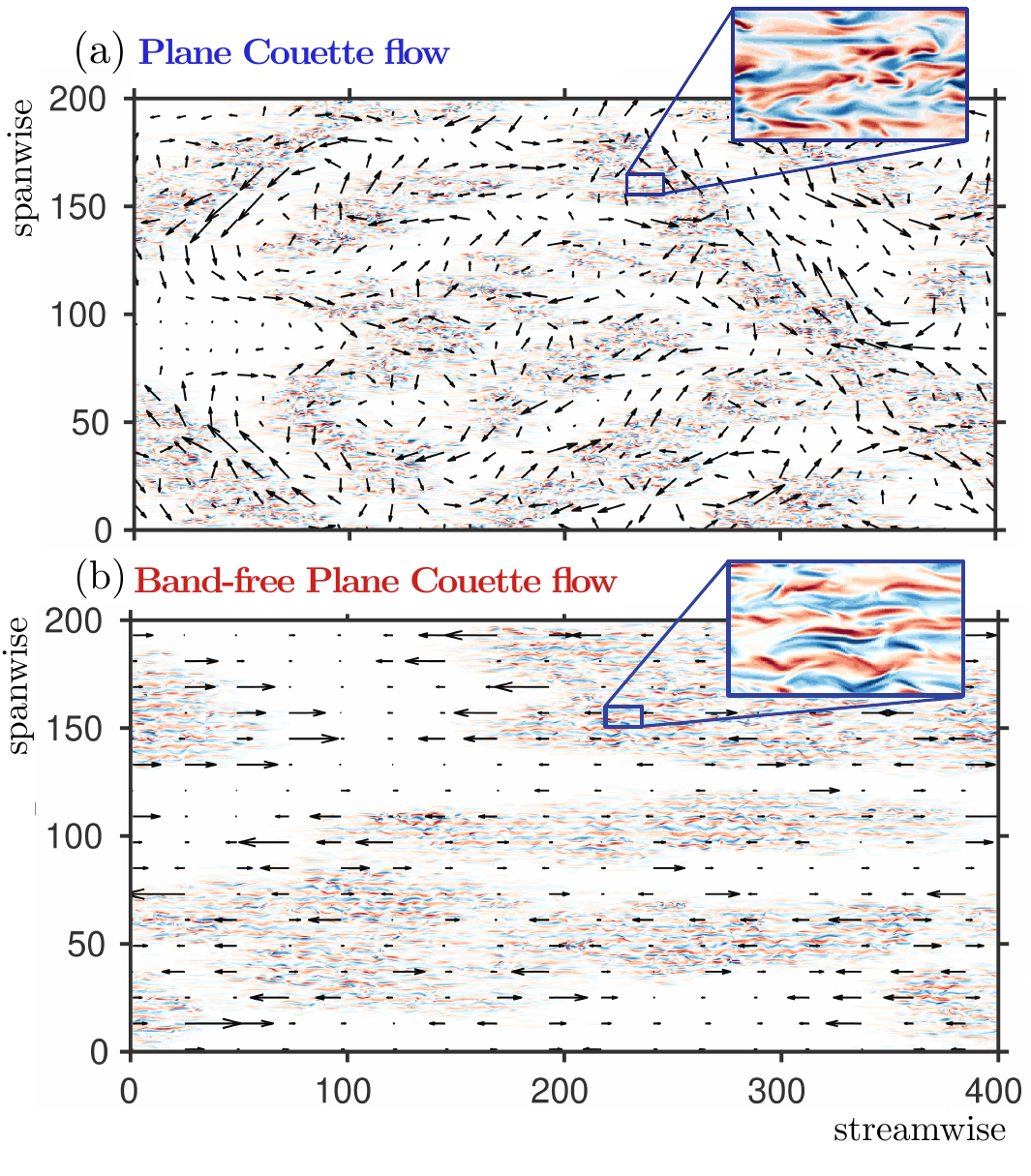}
\vspace{-3em}
\caption{Visualizations of (a) PCF and (b) band-free PCF at $Re=360$ in a domain of size $(L_\strm, L_\spwise)=(400,200)$.  Colors show the turbulent fluctuations, visualized via the wall-normal velocity $u_y$ 
at $y=0$ (blue: $ -0.2 \leq u_y < 0$, red: $0 < u_y \leq 0.2$). 
White ($u_y\simeq0$) signifies quiescent flow.
Arrows show the large-scale flow $(u_\strm,u_\spwise)$. The enlargements of $u_y$ in (a,b)} show the streakiness in the turbulent zones.
\label{fig:ld}
\end{figure}

The apparent simplicity of this story is belied by the multiple-scale structure of transitional turbulence typical in canonical wall-bounded shear flows.
At the scales of the wall separation, turbulence consists of streamwise vortices and streaks \cite{waleffe1997self}. 
At scales an order of magnitude larger, vortices and streaks organize into discrete coherent structures, known as puffs in straight pipe flow or oblique bands in planar flows \citep{prigent2002large,prigent2003long,barkley2005computational,duguet2010formation}; see Fig.~\ref{fig:ld}(a). 
To date, all experimental and numerical studies confirming universal DP scaling in wall-bounded flows have shown that bands or puffs control the percolation process \citep{lemoult2016directed, avila2011onset, chantry_universal, klotz2022phase,takeda2020intermittency, kohyama2022sidewall}.
This is at odds with the simpler scenario proposed by Pomeau based on front motion, metastability, and laminar gap formation.

Turbulent structures are symbiotically linked to large-scale mean flows \citep{coles1966progress,wygnanski1973transition,barkley2007mean,duguet2013oblique, couliou2015large, klotz2021experimental, marensi2023dynamics,van2009flow}
which dictate both their characteristic size and interactions
\citep{hof2010eliminating,samanta2011experimental,song2017speed, gome2}. Large-scale flow energizes turbulent structures \citep{song2017speed,gome1,van2009flow} and must be accounted for in theoretical treatments
\citep{barkley2011simplifying, barkley2016theoretical,
benavides2023model, barkley2011modeling, wang2022stochastic}.
Here, we realize a numerical experiment in which the large-scale flow in plane Couette flow (PCF) is controlled to eliminate the formation of oblique turbulent bands. The resulting flow will be called band-free PCF.
With this setup, we investigate the transition to turbulence in a hydrodynamic system 
without discrete large-scale structures and their associated mean-turbulent coupling.

We use the pseudo-spectral code {\tt Channelflow} \citep{channelflow} to 
carry out direct numerical simulations (DNS) of
the three-dimensional (3D) Navier-Stokes equations governing an incompressible viscous fluid between two parallel rigid plates moving at speeds $\pm U_{\rm wall}$. Velocities are non-dimensionalized by $U_{\rm wall}$, lengths by the half-gap $h$ between the plates. The Reynolds number is 
$Re=h U_{\rm wall}/\nu$, where $\nu$ is the kinematic viscosity.
Figure~\ref{fig:ld}(a) shows typical transitional flow containing large-scale oblique turbulent bands. 
The large-scale flow (arrows) is most pronounced at the interfaces separating turbulent and laminar regions \citep{prigent2003long, barkley2007mean, duguet2013oblique}.

The large-scale spanwise velocity is negligible for either fully laminar or fully turbulent flow, but appreciable along turbulent bands. 
By suppressing it, we seek to eliminate band formation.
We introduce streamwise (strm) and spanwise (span) large-scale cut-off wavenumbers $(K_\strm, K_\spwise)$.
For Fourier modes $|k_\strm|\leq K_\strm$ and $|k_\spwise| \leq K_\spwise$, we set the spanwise velocity to zero while retaining the usual momentum equations for streamwise and wall-normal velocity, including incompressibility.
This yields a two-component, three-dimensional (2C-3D) hydrodynamic system at large scales.
We apply the usual 3D Navier-Stokes equations at small scales, thereby preserving the mechanisms producing wall-bounded turbulence \citep{hamilton1995regeneration, waleffe1997self, liu2024lift}. 
The numerical procedure is given in the 
Appendix \citep{gomeSMarxiv}.

Figure \ref{fig:ld}(b) displays the flow computed with the cutoff window $(K_\strm, K_\spwise) = (0.24, 0.47)$, chosen so that the corresponding wavelengths $(\Lambda_\strm, \Lambda_\spwise) \simeq(26, 13)$ are smaller than the typical wavelengths $(\lambda_\strm,\lambda_\spwise) \simeq (100, 44)$ of the turbulent bands in PCF \citep{prigent2003long}. The large-scale flow is streamwise-oriented by construction.  Our procedure 
has the desired effect of eliminating the turbulent bands -- the laminar-turbulent interfaces do not have the well-defined angles and widths seen in Fig.~\ref{fig:ld}(a).

In order to study spatio-temporal dynamics, we perform simulations in a long slender domain tilted with respect to the streamwise direction by $\theta=24^\circ$, a typical angle at which bands occur; see \citep{prigent2003long,barkley2005computational}.
We denote the slender direction parallel to the bands by $x$ and the long direction perpendicular to them by $z$.
This geometry has been shown 
to capture important features of turbulent bands while reducing their behavior to quasi-1D dynamics along $z$ \citep{barkley2005computational, barkley2007mean, shi, lemoult2016directed, gome2020statistical}. For consistency and comparison, we use the same domain for band-free PCF, in which we suppress large-scale {\em spanwise} velocity.
(See Appendix \citep{gomeSMarxiv}.)


\paragraph*{Mean-turbulent interaction \textemdash}

\begin{figure} 
\includegraphics[width=1\columnwidth]{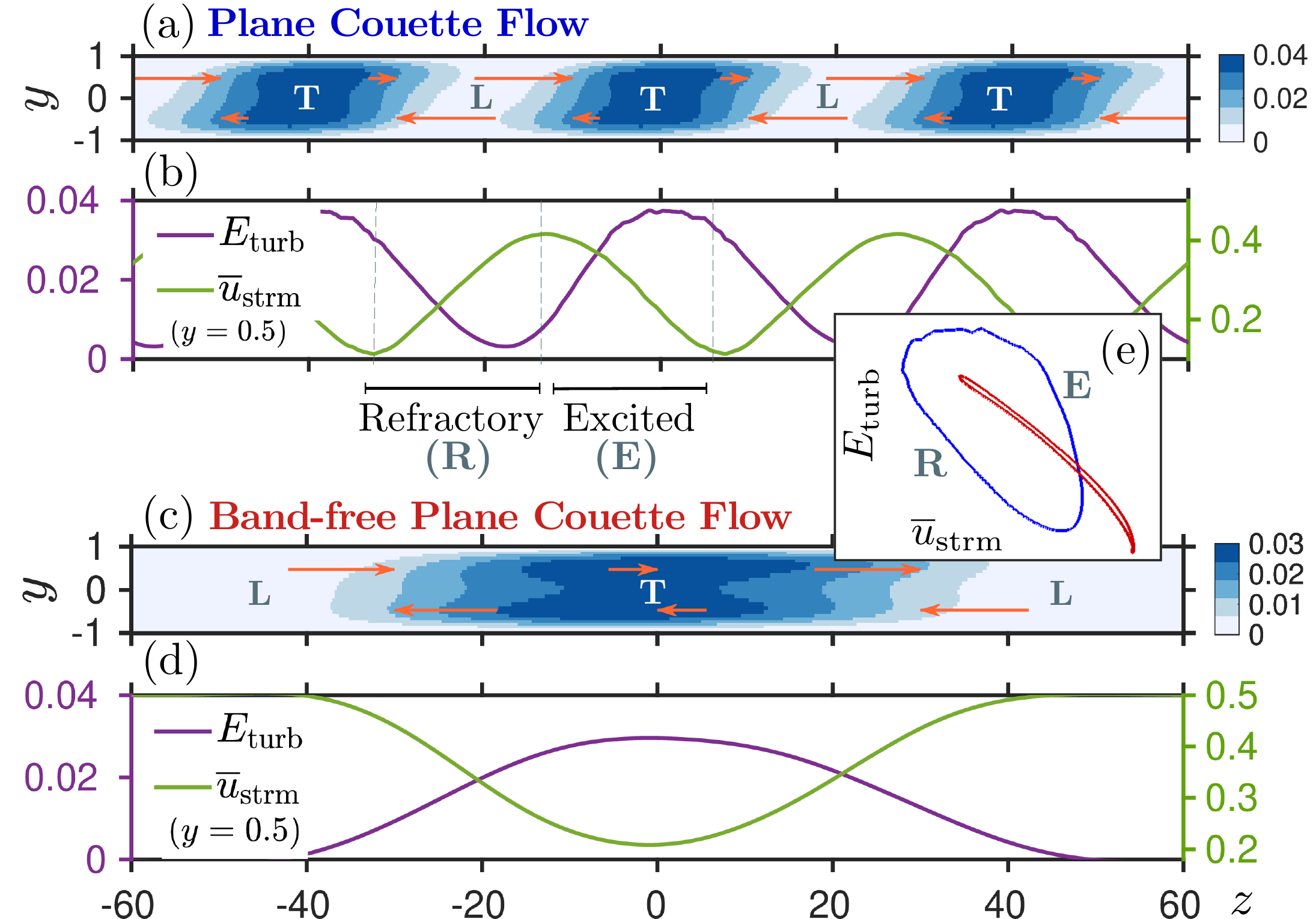}
\vspace{-2em}
\caption{
Turbulent kinetic energy $E_{\rm turb}$ and mean streamwise velocity $\overline{u}_{\strm}$ in (a,b) PCF and (c,d) band-free PCF.
Contour plots (a,c) show 
$E_{\rm turb}(y,z)$ and arrows show $\overline{u}_{\strm}(\pm 0.5,z)$
Unlike in (a), turbulent zones in (c) do not have a selected width.
Curves in (b,d) show $E_{\rm turb}$ and $\overline{u}_{\strm}$ as a function of $z$ at $y=0.5$. (e) Phase-plane representation of 
$\overline{u}_{\strm}$ and $ E_{\rm turb}$ 
at $y=0.5$ in PCF (blue) and band-free PCF (red). 
In contrast to PCF, turbulence and mean-flow in band-free PCF are in phase.
Abbreviations: L, laminar; T, turbulent; R, refractory; E, excited. 
}
\label{fig:mf}
\end{figure}

Prior to comparing the transition scenarios in PCF and band-free PCF, we focus on the coupling between turbulence and large-scale mean flow.
We decompose the velocity $\utot = \ubar+\uprime$, where averages $\overline{(\cdot)}$ 
are taken over $x$ and time periods during which turbulent fronts are approximately stationary.
The turbulent kinetic energy $E_{\rm turb} \equiv \frac{1}{2} \overline{\uprime \cdot \uprime}$ and the mean streamwise velocity $\overline{u}_{\rm strm}$ are plotted in Fig.~\ref{fig:mf}.

The significant difference between PCF and band-free PCF is in the spatial phase relation between $E_{\rm turb}$ and $\overline{u}_{\rm strm}$, seen at $y=0.5$ in Figs.~\ref{fig:mf}(b,d).
(Reflected plots would be obtained at $y=-0.5$.)
Turbulence extracts energy from the local mean shear,
thereby flattening the profile and reducing $\overline{u}_{\rm strm}$ at $y=0.5$.
For bands in PCF, advection by the large-scale flow (arrows) redistributes momentum and energy from laminar to turbulent regions 
\citep{barkley2007mean} (see especially Fig.~6(b) in \citep{gome1}).
It is the competing mechanisms of mean-flow flattening by turbulence and mean flow fuelling nearby turbulence
that are encrypted in the phase relation between $\overline{u}_\strm$ and $E_{\rm turb}$ (see Fig.~\ref{fig:mf}(e)).
When turbulence is \emph{excited}  
($E_{\rm turb}$ increases, see Fig.~\ref{fig:mf}(b)), it is first 
fuelled by the large-scale flow.
Hence, $\overline{u}_{\rm strm}$ does not react directly to it and
decreases only after a phase shift, 
and then vice versa in the \emph{refractory} region.
Excited and refractory regions are essential for sustaining 
localised turbulent regions and for their duplication and interactions
\citep{van2009flow, hof2010eliminating, samanta2011experimental, barkley2016theoretical}.

In band-free PCF, we find that turbulence and mean flow are not phase shifted and excited and refractory zones are absent, i.e.\ $E_{\rm turb}$ is a single-valued function of $\overline{u}_\strm$ (Fig.~\ref{fig:mf}(e)). 
This follows from
the approximate $z$-reflection symmetry apparent in Fig.~\ref{fig:mf}(c,d). (See Supplemental Material, \S I.D).
This has major consequences for the way that turbulence emerges in this flow.


\paragraph*{Transition with and without bands.\textemdash}

\begin{figure}
\includegraphics[width=1\columnwidth]{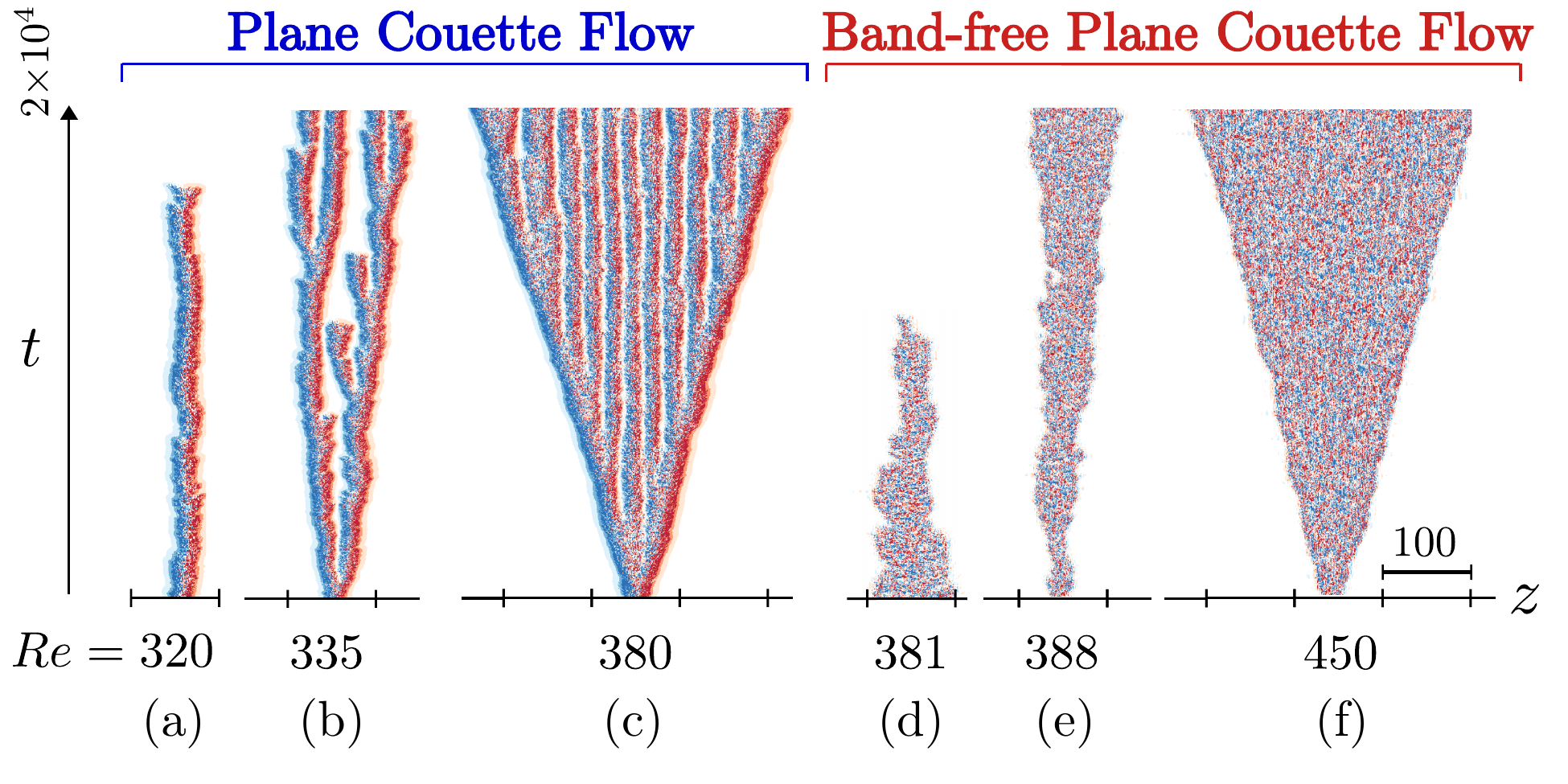}
\vspace{-2em}
\caption{Band and slug regimes in (a,b,c) PCF and (d,e,f) band-free PCF initiated from a turbulent patch. Colors show the spanwise velocity $u_\spwise$ at $y=0$, $x=0$ as a function of $z$ and $t$
blue: $-0.1 \leq u_\spwise \leq 0$, red:  $0 \leq u_\spwise \leq 0.1$).
The wide blue and red areas visible in PCF result from the large-scale flow.}
\label{fig:spacetime}
\end{figure}

Transition scenarios in PCF and in band-free PCF are illustrated by spatio-temporal diagrams in Fig.~\ref{fig:spacetime}.  Each simulation is initiated with a localized turbulent patch.

\begin{figure}
\includegraphics[width=\columnwidth]{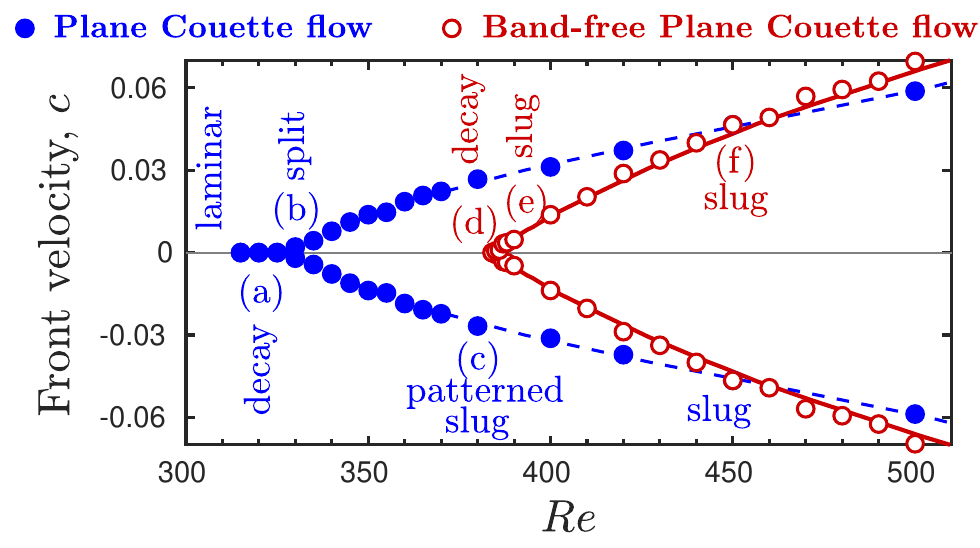}
\vspace{-2.5em}
\caption{Front
velocity as a function of $Re$ in both PCF and band-free PCF. 
Labels (a-f) correspond to the visualizations of Fig.~\ref{fig:spacetime}(a-f).
The points of zero propagation speed are $Re_0 \simeq 325$ for PCF \citep{shi} and $Re_0 \simeq 383$ for band-free PCF.
Speeds below $Re_0$ in band-free PCF are not shown; see Supplemental Material.
The blue curve guides the eye, while the continuous red curves come from simulations of model \eqref{eq:DP_model}.
}
\label{fig:slug}
\end{figure}

The scenario for PCF is well documented in this geometry \citep{barkley2005computational, shi,lemoult2016directed, gome2}.  
Below $Re \simeq 350$, turbulent bands are localized metastable structures that survive for long times before decaying (Fig.~\ref{fig:spacetime}(a)) or proliferating by splitting (Fig.~\ref{fig:spacetime}(b)) \citep{shi}.
Decay and splitting are memoryless processes, in that their associated waiting times are exponentially distributed. Above $Re_0 \simeq 325$
\citep{shi}, the proliferation of a single band becomes more probable than its decay, and turbulence survives, albeit in intermittent form \citep{lemoult2016directed, klotz2022phase}.
In the thermodynamic limit of large systems with many bands and long times, this intermittency gives rise to a critical point in the DP universality class at the estimated critical value $Re_c =328.7$ \citep{lemoult2016directed, klotz2022phase}.
($Re_c$ and $Re_0$ are close, but distinctly different: 
$Re_0$ is determined from single isolated bands while $Re_c$ is influenced by interactions between bands.)
For $Re \gtrsim 350$, e.g. Fig.~\ref{fig:spacetime}(c), turbulence proliferates via expanding fronts moving at equal and opposite speeds (\emph{slug} phase \citep{barkley2016theoretical}). 
The mean left-going and right-going front propagation speeds are shown in Fig.~\ref{fig:slug}.  
Below $Re\simeq 450$, the fronts delimit a nearly periodic interior pattern (Fig.~\ref{fig:spacetime}(c)) \citep{shi}, which is absent above $Re\simeq 450$ (not shown).

By design, band-free PCF does not exhibit 
discrete turbulent structures or patterns. Turbulence contracts and expands, not via decay and splitting of discrete structures, but rather via fluctuating front motion, shown in Figs.~\ref{fig:spacetime}(d)--(f), whose expansion speeds are also shown in Fig.~\ref{fig:slug}.
Importantly, the decay of turbulence is not a memoryless process because the mean lifetime of a patch depends on its initial size, unlike in PCF and pipe flow
\citep{shi, avila2010transient, avila2011onset}, but like low-$Re$ plane channel flow \citep{mukund2021aging,xu2022size}.

\begin{figure}[t]
 \includegraphics[width=\columnwidth]{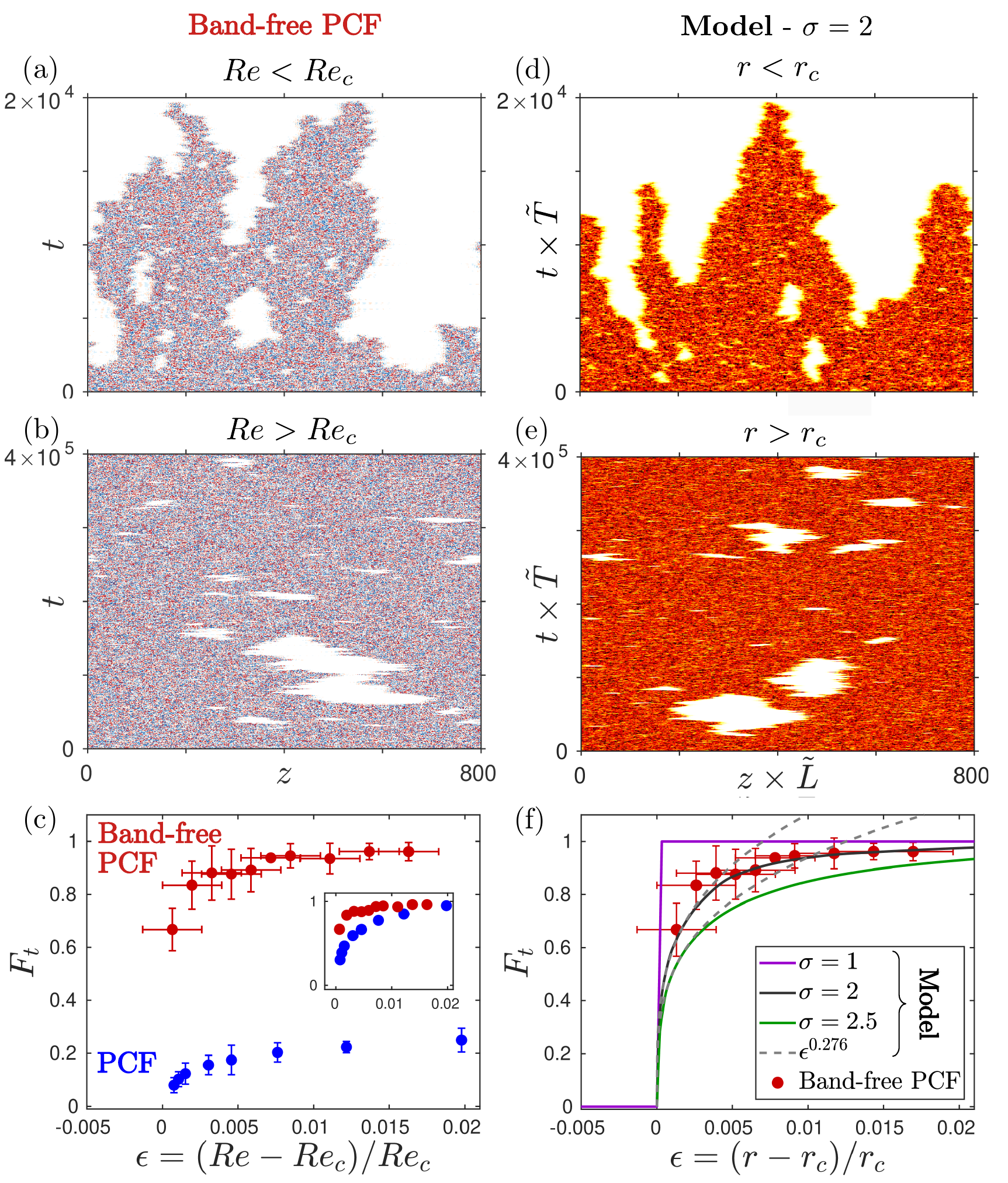}
\caption{Laminar-turbulent percolation in 
band-free PCF at (a) $Re=370 < Re_c$ and (b) $Re=385>Re_c$.
(c) Equilibrium turbulent fraction $F_t$ as a function of $\epsilon \equiv (Re-Re_c)/Re_c$  in PCF (blue, data from Ref.~\citep{lemoult2016directed})
and band-free PCF (red). 
The inset in (c) shows PCF data rescaled by $F_t(\epsilon=0.02)$ and confirms the more abrupt transition in band-free PCF.
(d, e) Space-time visualisation of model \eqref{eq:DP_model} with $\sigma=2.0$ at 
(d) $r<r_c \simeq 0.2517$ and (e) $r>r_c$.
Time and length in the model are rescaled with $\tilde{T}$ and $\tilde{L}$, and $r$-values are chosen such that $r-r_0=b(Re-Re_0)$.
(f) Same as (c) for model \eqref{eq:DP_model} with $\sigma=1.0$ (first-order), $2.0$ and $2.5$ (both second-order cases exhibiting DP scaling $\epsilon^{0.276}$, see dashed lines).}
\label{fig:model_vs_dns}
\end{figure}

Figure \ref{fig:model_vs_dns}(a,b) shows the spatio-temporal dynamics from a fully turbulent initial state in a long domain
below ($Re = 370$) and above ($Re=385$) the critical point $Re_c$ in band-free PCF.
Laminar gaps nucleate within the turbulent flow and interfaces fluctuate.
Below $Re_c$, turbulence predominantly contracts and asymptotically the flow is laminar.
Above $Re_c$, laminar gaps are created but eventually close as a consequence of preferred turbulent expansion.

The order parameter for the transition is the equilibrium turbulent fraction, $F_t$, 
which is the mean proportion of turbulent flow at statistical equilibrium (see Supplemental Material).
$F_t$ is plotted as a function of $Re$ in Fig.~\ref{fig:model_vs_dns}(c) for both PCF \citep{lemoult2016directed} and band-free PCF. 
For comparison, we use reduced Reynolds numbers, $\epsilon \equiv (Re-Re_c)/Re_c$.
For PCF, $Re_c \simeq 328.7$ \citep{lemoult2016directed}. For band-free PCF we estimate $Re_c\simeq 383.5$.
Without bands, the transition to uniform turbulence occurs over a significantly shorter range of $Re$ 
($F_t\simeq 0.9$ at $\epsilon \simeq 0.01$, while $F_t\simeq 0.2$ at $\epsilon \simeq 0.01$ in PCF). 

Measuring $F_t$ near $\epsilon = 0$ is exceedingly costly because scales diverge and hence simulations become susceptible to finite-size, finite-time effects. We have not attempted to determine critical scalings associated with transition in band-free PCF.

\paragraph*{A model for percolation via front motion.\textemdash}
\label{sec:model}
Published models for puffs and bands contain at least two fields \citep{barkley2011simplifying,barkley2016theoretical,barkley2011modeling,wang2022stochastic,benavides2023model}; two are necessary to capture the interaction between mean-flow and turbulence responsible for discrete structures; 
see Fig.~\ref{fig:mf}(a,b).
Band-free PCF lacks discrete structures and mean-flow and turbulence are
slaved, see Fig.~\ref{fig:mf}(c,d). This suggests modelling band-free PCF with a single scalar field $q(z,t)$ representing 
the local turbulent energy.

Following \citep{pomeau, pomeau2015transition, barkley2016theoretical}, we consider the stochastic model
\begin{align}
\partial_t q &=  \partial^2_{z} q  - \partial_q V + \sigma q \xi
    \label{eq:DP_model}\\
\intertext{with the double-well potential}    
V (q) &\equiv \frac{q^2}{2} ~ \left[ 1 + (r+ 1) \left(\frac{q^2}{2} - \frac{4q}{3}\right) \right]\nonumber,
\end{align}
$\xi$ a space-time Gaussian white noise of unit variance and $\sigma$ a parameter controlling the noise strength.
Parameter $r$ plays the role of $Re$. 
$V$ has one local minimum at $q=0$ (laminar) and one at $q>0$ (turbulent state), the latter being the lowest minimum for $r>1/8$.
Multiplicative noise $\sigma q \xi$ represents turbulent fluctuations and vanishes at $q=0$, making it an absorbing state; 
see \citep{munoz1998nature, munoz2003stochastic, munoz2003multiplicative}.
Hence model~\eqref{eq:DP_model} describes fronts between an absorbing
and a fluctuating state, 
just as Pomeau originally envisioned for the transition to turbulence
\cite{pomeau,pomeau2015transition,barkley2016theoretical}.

We solve~\eqref{eq:DP_model} via a finite-difference scheme under the It\^o representation \citep{gardiner1985handbook}. 
We first simulate expanding slugs.
Front velocities are found to closely obey the self-similar expression $c \simeq F(r-r_0(\sigma))$ 
where $F(0)=0$ and $r_0 (\sigma) \simeq 1/8 + 3 \sigma^2/100$. 
$F$ is independent of $\sigma$. Front speeds in band-free PCF,
shown in Fig.~\ref{fig:slug}, follow $c_{\rm DNS} \simeq a F(b(Re-Re_0))$ (solid red line), where $a$ and $b$ rescale velocities and $Re$, respectively.
The agreement is excellent.

The fit of the model speeds to those of band-free PCF
establishes the correspondence between $r$ and $Re$.
We determine the ratio $\tilde{L}\simeq 3.0$ of band-free PCF
to model length scales from the width of laminar-turbulent fronts.
From $\tilde{L}$ and $a$, we obtain a ratio of time scales between  
band-free PCF 
and model, $\tilde{T}= \tilde{L}/a\simeq 16$ (see Supplemental Material).
The final model parameter $\sigma$ is chosen so as to reproduce the phenomenology of 
band-free PCF as we now show.

With $\sigma=2.0$,
simulations of \eqref{eq:DP_model} with space, time and $Re$ rescaling 
are visualized in Fig.~\ref{fig:model_vs_dns}(d, e).  
The model closely reproduces the dynamics of fluctuating fronts and 
the nucleation of laminar gaps within turbulent flow as seen
in band-free PCF (Fig.~\ref{fig:model_vs_dns}(a, b)).

The noise intensity $\sigma$ indirectly controls the rate of laminar-gap nucleation, and, consequently, the order of the phase transition, as we show 
in Fig.~\ref{fig:model_vs_dns}(f).
With sufficiently strong noise ($\sigma=2.0$ and $2.5$), the transition is continuous and exhibits the scalings of DP 
($F_t \sim \epsilon^{{0.276}}$, spatial and temporal correlations $\xi_\perp \sim \epsilon^{1.097}$ and $\xi_{\parallel} \sim \epsilon^{1.734}$; see Supplemental Material).
Meanwhile, at low noise ($\sigma=1.0$) the transition is discontinuous, with no intermediate $0< F_t < 1$ sustained.
The transition becomes sharper as $\sigma$ is decreased.

We find that the value $\sigma= 2.0$ captures the dependence of turbulent fraction on $Re$ in band-free PCF (Fig.~\ref{fig:model_vs_dns}(f)) with similar sharp increase in $F_t$, and specifically gives $F_t (\epsilon \simeq 0.01) \simeq 0.9$.
(The accuracy of this match is limited by the precision of $Re_c$.)
Because this value of $\sigma$ corresponds to a continuous phase transition, this strongly suggests that band-free PCF, like PCF, undergoes a continuous transition 
in the DP universality class.

\paragraph*{Conclusion and discussion.\textemdash}

In subcritical shear flows, turbulence appears, not by increasing its intensity, but by occupying an increasing proportion of space as Reynolds number is increased.
In previous studies of canonical straight pipe or planar flows, transition occurs via the percolation of discrete turbulent structures 
that individually decay or self-replicate. Large-scale mean flow plays a crucial role in these systems by selecting these discrete structures. 

Here, we investigate a plane-Couette setup without discrete structures.
Transition follows a distinctly different scenario, mediated by expanding turbulent fronts and spontaneous nucleation of laminar gaps.
The observed dynamics are very well captured by a simple stochastic, double-well model with a single field. 
The order of this phase transition is governed by the nucleation rate of laminar gaps, which depends on the level of fluctuations. 

This front-moving transition should be pervasive in systems where discrete puffs or bands are absent. Numerous flows, such as 
bent pipes \citep{rinaldi2019vanishing}, body-forced pipes \citep{zhuang2023discontinuous}, stably-stratified flows \citep{rorai2014turbulence}, suction boundary layer \citep{khapko2016turbulence} or 
constrained Couette flow \citep{duguet2011stochastic, pershin2019dynamics}, lack discrete turbulent structures and may follow this scenario.
Our Letter suggests that in such flows, 
the level of turbulent fluctuations plays a crucial role in dictating whether front-moving transition is first-order, 
as recently reported in experiments of curved and body-forced pipes \citep{zhuang2023discontinuous}, or second-order and within the DP universality class.

\begin{acknowledgments}

We thank Anna Frishman, Yohann Duguet, 
Santiago Benavides, François P\'etr\'elis and Tobias Grafke for fruitful discussions. The calculations for this work were performed using high performance computing resources provided by the Grand Equipement National de Calcul Intensif at the Institut du D\'eveloppement et des Ressources en Informatique Scientifique (IDRIS, CNRS) through Grant No.~A0142A01119. This work was partly supported by a grant from the Simons Foundation (Grant No.~662985, D.B. and L.S.T.).
\end{acknowledgments}


\newpage
\onecolumngrid
\appendix*


\bigskip
\begin{center}
   {\bf \large Appendix: Supplementary Material for \\[0.2cm]
   
Phase transition to turbulence via moving fronts}
\end{center}

\subsection{Suppressing large-scale spanwise flow in PCF}
\label{sec:method}

Large-scale spanwise velocity is an emergent feature of bands in transitional shear flows. 
We have introduced a numerical procedure which eliminates band formation in plane Couette flow (PCF) by suppressing large-scale spanwise velocity. Meanwhile, we keep large-scale variations of the streamwise and wall-normal velocities, so that laminar-turbulent intermittency is preserved.

For small-scale modes, i.e. for $\boldsymbol{k}$ 
such that $|k_\strm| > K_\strm$ or $|k_\spwise| > K_\spwise$, 
we solve the usual 3D Navier-Stokes equations:
\begin{subequations}
    \begin{align}
    \frac{\pd \ujhat}{\pd t} + \widehat{\nabla}_j \whp
    &= - \widehat{N}_j + \frac{1}{Re} \widehat{\nabla}^2 \ujhat \text{ for } j = \strm, y, \spwise\\
      \widehat{\nabla} \cdot \uhat &= 0  
      \label{eq:3D_NS_c}
    \end{align}
    \label{eq:3D_NS}%
\end{subequations}
where $\widehat{N}_j \equiv \sum_{\boldsymbol{p}+\boldsymbol{q} = \bk} \widehat{\partial_l u_l} (\boldsymbol{p}) \widehat{u_j} (\boldsymbol{q})$ are the nonlinear advective terms.

For the large-scale modes, i.e. $\boldsymbol{k}$ such that $|k_\strm|\leq K_\strm$ and $|k_\spwise|\leq K_\spwise$, we solve the following system
for the Fourier coefficient $\uhat (k_\strm, k_\spwise)$
\begin{subequations}
    \begin{align}
    \frac{\pd \ujhat}{\pd t} + \widehat{\nabla}_j \whp
    &= - \widehat{N}_j + \frac{1}{Re} \widehat{\nabla}^2 \ujhat \text{ for } j = \strm, y\label{eq:2D_NS_a}\\     \widehat{\nabla}_{2D} \cdot \uhat &= 0  \\
      \whu_\spwise &= 0
      \label{eq:2D_NS_c}
    \end{align}
    \label{eq:2D_NS}%
\end{subequations}
where $\widehat{\nabla}_{2D} \equiv (i k_\strm, \pd_y, 0)$.
This is a two-component, three-dimensional (2C-3D) system for these low-wavenumber modes. 

\subsection{Numerical setup}
\label{sec:numerical}
We solve the 
3D Navier-Stokes equations 
using {\tt Channelflow} \citep{channelflow}, which uses a semi-implicit pseudospectral 
in primitive variables, decomposed in Fourier-Chebychev modes. 
We briefly
explain the underlying numerical principles, derived from \citet{canuto2007spectral}. 

Equations \eqref{eq:3D_NS} are discretized in time and 
in the Fourier directions as follows:
\begin{subequations}\begin{align}
    (\pd_y^2  - \lambda) \whu_{\strm}(t) &=   \whR_\strm - i k_\strm  \whp, \label{eq:ux}
    & \text{    with } \whu_\strm(\pm1)=0 \\
    (\pd_y^2  - \lambda) \whu_y(t) &=   \whR_y - \pd_y  \whp, \label{eq:uy}
    & \text{    with } \whu_y(\pm1)=0 \\
     (\pd_y^2  - \lambda) \whu_\spwise(t) &=   \whR_\spwise - i k_\spwise  \whp 
     & \text{    with } \whu_\spwise(\pm1)=0 \label{eq:uz}\\
     ik_\strm \whu_\strm + \pd_y \whu_y + ik_\spwise \whu_\spwise &= 0 
     \label{eq:div}
\end{align}\end{subequations}
The terms $\widehat{ \boldsymbol{R}}$ come from explicit integration of the non-linear term $\widehat{\boldsymbol{N}}$ and time derivative and are given by 
\begin{align}
   \widehat{ \boldsymbol{R} } &\equiv \sum_{s} \frac{\alpha_s}{\dt} \uhat(s)   - \beta_s \widehat{\boldsymbol{N}}(s),
   \label{eq:RHS}
\end{align}
where $s$ ranges over previous timesteps, 
$\dt$ is the numerical timestep, and
$\alpha_s$ and $\beta_s$ are coefficients that depend on the time-stepping method. 
Equation \eqref{eq:div} is eliminated in the usual way by acting with the divergence $(ik_\strm,\pd_y,ik_\spwise)$ on \eqref{eq:ux}-\eqref{eq:uz}, leading to the pressure Poisson equation, and by substituting the boundary conditions on $\whu_\strm$, $\whu_\spwise$ into \eqref{eq:div}. 
This yields the following system for $(\whu_y,\whp)$ for each Fourier wavenumber pair $(k_\strm,k_\spwise)$:
\begin{align}
  \begin{drcases}
        (\pd_y^2  - k_\strm^2 - k_\spwise^2) \:\whp(t) &= i k_\strm \whR_\strm + \pd_y \whR_y+ i k_\spwise \whR_\spwise
        \\
        (\pd_y^2  - \lambda) \:\whu_y (t) &=   \whR_y -  \pd_yp
   \end{drcases}
   & \text{      with } 
   \begin{dcases}
   \pd_y\whu_y(\pm1)&= 0\\ 
   ~~~\whu_y(\pm1)&=0 
   \end{dcases}
   \label{eq:usual}
\end{align}
where $\lambda \equiv 2/\dt + (k_x^2 + k_z^2)/Re$ arises from the implicit treatment of the viscous term.
This system of two Helmholtz problems, coupled by their four boundary conditions on $\hat{u}_y$,
is solved via an influence-matrix method. 
Equations \eqref{eq:ux} and \eqref{eq:uz} for the two remaining components $\whu_\strm, \whu_\spwise$
are uncoupled and easily solved.

We now turn to the suppression of spanwise flow in large-scale modes with $|k_\strm| \leq K_\strm$ and $|k_\spwise| \leq K_\spwise$.
Substituting $\whu_\spwise=0$ for \eqref{eq:uz} and carrying out the same operations as before to eliminate \eqref{eq:div} now leads to
\begin{align}
  \begin{drcases}
        (\pd_y^2  - k_\strm^2) \:\whp(t) &= i k_\strm \whR_\strm + \pd_y \whR_y 
        \\
        (\pd_y^2  - \lambda) \:\whu_y (t) &=   \whR_y -  \pd_yp
   \end{drcases}
   & \text{      with } 
   \begin{dcases}
   \pd_y\whu_y(\pm1)&= 0\\ 
   ~~~\whu_y(\pm1)&=0 
   \end{dcases}
   \label{eq:unusual}
\end{align} 
The influence matrix must be modified accordingly to solve 
system \eqref{eq:unusual}.
We note that a related (though different) strategy was used by \citet{jimenez1999autonomous, jimenez2022streak}, so as to study the cyclic mechanisms in the turbulence production in wall-bounded flows.

\subsection{Simulations of band-free PCF in a tilted domain}
\label{app:tilted}

We now define the tilted coordinate system $(x,z)$ via
\begin{subequations}
\label{tilted}
\begin{align}
\be_\strm &= \quad\cos{\theta} \, \be_{\xt} + \sin{\theta} \, \be_{\zt} = \;\;\alpha\be_{\xt} + \beta\be_{\zt}\\
\be_\spwise &= -\sin{\theta } \, \be_{\xt} + \cos{\theta} \, \be_{\zt} = -\beta \be_{\xt} + \alpha \be_{\zt},
\end{align}
\end{subequations}  
illustrated in Figure \ref{fig:tilted}. 
As with prior studies \cite{barkley2005computational}, the size $L_x$ is chosen to $L_x=10$ so that bands are homogeneous in this direction. In this geometry, we will suppress spanwise flow for modes $k_x=0$ and $|k_z|<K_z$, where $K_z$ denotes a cutoff wavenumber in the $z$ direction. Note that while the geometry is tilted with respect to the streamwise-spanwise coordinates, we still suppress spanwise flow.
To set $\whu_\spwise=0$ while preserving the evolution equations for $\whu_\strm$ and $\whu_y$ we have
\begin{subequations}\begin{align}
    (\pd_y^2  - \lambda) \whu_\strm &=   \whR_\strm - i k_\strm  \whp, \label{eq:tux}
    & \text{    with } \whu_\strm(\pm1)=0 \\
    (\pd_y^2  - \lambda) \whu_y &=   \whR_y - \pd_y  \whp, \label{eq:tuy}
    & \text{    with } \whu_y(\pm1)=0 \\
     \whu_\spwise &=   0 
     & \text{    with } \whu_\spwise(\pm1)=0 \label{eq:tuz}\\
     ik_\strm \whu_\strm + \pd_y \whu_y &=0
     \label{eq:tdiv}
\end{align}\label{eq:tu}\end{subequations}
where $k_\strm = \alpha k_{\xt} + \beta k_{\zt}$ and $\whR_\strm = \alpha \whR_{\xt} + \beta \whR_{\zt}$.
Taking the 2D divergence $(ik_\strm,\pd_y)$ of \eqref{eq:tux}-\eqref{eq:tuy} and substituting \eqref{eq:tdiv} leads to 
\begin{align}
  \begin{drcases}
(\pd_y^2-k_\strm^2) \whp    &= ik_\strm \whR_\strm +\pd_y \whR_y \\
    (\pd_y^2  - \lambda) \whu_y &=   \whR_y - \pd_y  \whp,
\end{drcases}
   & \text{      with } 
   \begin{dcases}
   \pd_y\whu_y(\pm1)&= 0\\ 
   ~~~\whu_y(\pm1)&=0, 
   \end{dcases}
\end{align}
which is identical to the non-tilted case \eqref{eq:unusual}.
For the remaining components, we use:
\begin{subequations}\begin{align}
0 = \whu_\spwise &= -\beta u_{\xt} + \alpha u_{\zt} \Longrightarrow \whu_{\zt} = \beta \whu_{\xt}/\alpha \label{eq:suba}\\
\whu_\strm&=\alpha \whu_{\xt} + \beta \whu_{\zt} = \alpha \whu_{\xt} + \beta^2\whu_{\xt}/\alpha
= (\alpha^2 +\beta^2)\whu_{\xt}/\alpha  = \whu_{\xt}/\alpha \label{eq:subb}
\end{align}\end{subequations}
leading to the following equations for $\whu_{\xt}$ and $\whu_{\zt}$:
\begin{subequations}\begin{align}
    (\pd_y^2  - \lambda) \frac{1}{\alpha} \whu_{\xt} &=   (\alpha \whR_{\xt} + \beta \whR_{\zt}) -i(\alpha k_{\xt} + \beta k_{\zt}) \whp, \label{eq:tux3}
    & \mbox{with } \whu_{\xt}(\pm1)=0 \\
\whu_{\zt} &= \beta\whu_{\xt} /\alpha
     & \mbox{with } \whu_{\zt}(\pm1)=0 \label{eq:tuz3}\
\end{align}\end{subequations}

\subsection{Spectra in PCF and band-free PCF and choice of cutoff}
\label{subsec:remarks}

One can view the 
suppression of the large-scale spanwise velocity as an effective large-scale forcing on the spanwise momentum equation, that counteracts the effects of the nonlinear term $N_\spwise$ and the pressure $\partial_\spwise p$. 
As a consequence, our procedure
does not preserve continuity in Fourier space of the velocity components or of $\boldsymbol{N}$.

\begin{figure}
    \centering
\includegraphics[width=1\columnwidth]{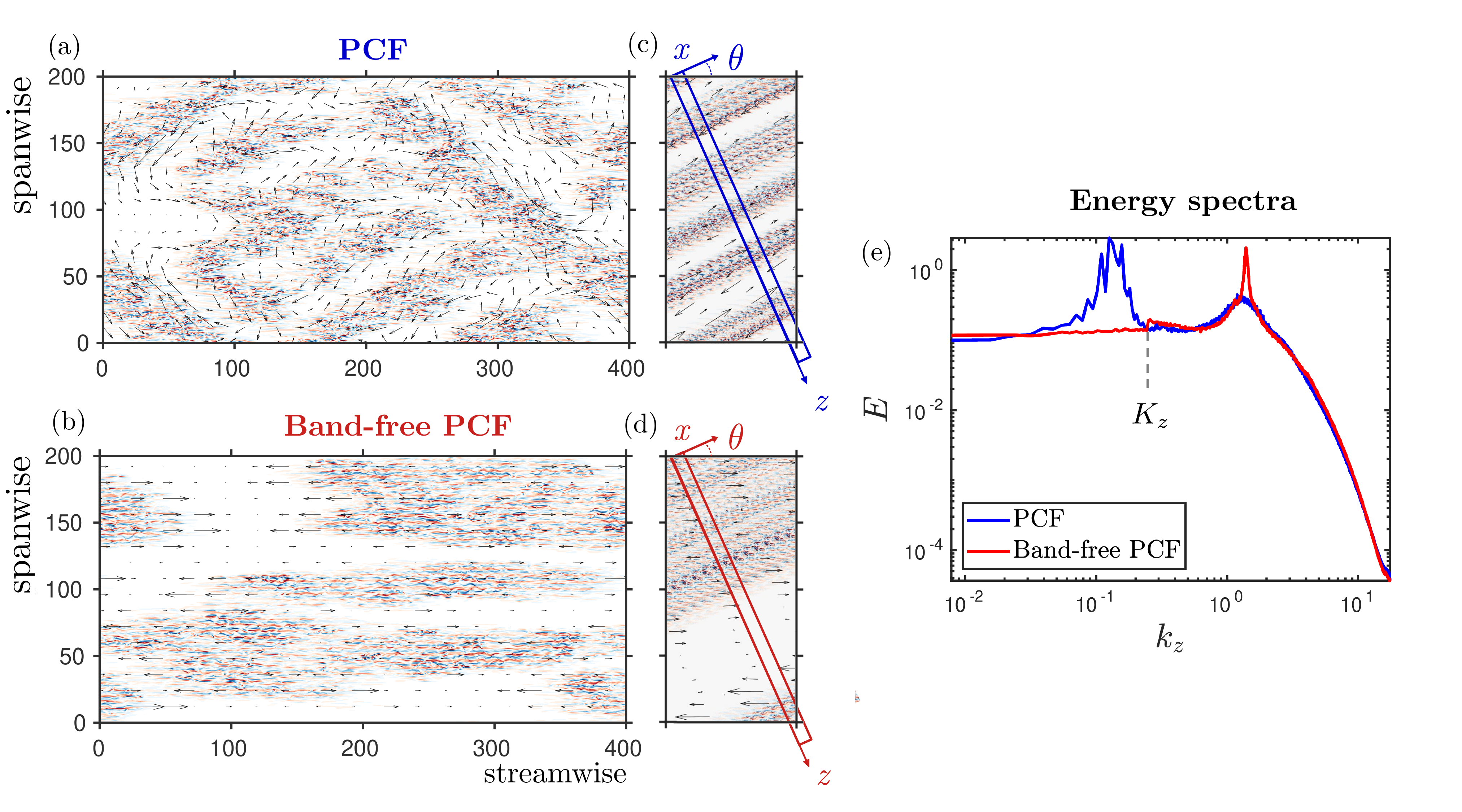} 
\caption{Visualization of (a) PCF, (b) band-free PCF in a streamwise-spanwise oriented domain, and (c,d) in a quasi-1D tilted domain. (e) Spectral energy $E (k_{z}) \equiv \frac{1}{2}\left< \uhat \cdot \uhat \right>_{x, t}$. Average is taken over $(x, t)$ in the tilted domain. Flow and spectra are shown at $y=0$, $Re=360$. Cutoff wavenumber $K_z = 0.24$ is indicated.  
}
    \label{fig:tilted}
\end{figure}
Figure \ref{fig:tilted}(e) shows the energy spectrum in a tilted domain of both PCF and band-free PCF. We compute $E (k_{z}) \equiv \frac{1}{2}\left< \uhat \cdot \uhat \right>_{x, t}$, with $\uhat$ the $z$-Fourier transform of total velocity $\utot$, and where the average is carried out over $x$ and $t$, at $Re=360$.
The characteristic large-scale motions in PCF associated with the bands ($k_{\rm bands}\simeq 0.14$) are damped to an almost constant energy spectrum in the band-free case. 
As expected, the energy 
spectrum shows a small discontinuity at the cutoff wavenumber $K_{\zt}$
(here $K_{\zt}=0.24$).

Furthermore, the energy at small scales is increased as a result of band suppression, as can be seen in
an increase in the small-scale energy peak corresponding to streaks and rolls \cite{gome1}. 
The band-suppression 
method that we use is indeed not designed to conserve the total dissipation in the flow, nor the dissipation at small scales; hence the energy that originally feeds bands is redistributed to the streaks in an uncontrolled way.

We chose a cutoff wavenumber $K_z$ within the range $0.20 < K_z < 0.8$. 
If $K_z$ is chosen below $0.20$, it is too close to $k_{\rm bands}\simeq 0.14$ and the flow still sustains discrete bands with selected size close to $2\pi/K_z$. 
For a range of $K_z$ above $0.20$ discrete structures are absent and we observe that transition takes place via moving fronts.
However, the value of $K_z$ controls the dissipation within the flow, hence the value of critical Reynolds number $Re_c$: 
the higher $K_z$, the more streaks and rolls are energized, hence the lower $Re_c$.
We have not explored $K_z> 0.8$, as $K_z$ would be close to the spacing of streaks and rolls ($k_z\sim 1$), and could affect the self-sustaining process.
In the main text, we show results at $K_z=0.24$.
Note that in the untilted, streamwise-spanwise configuration the choice of cutoff wavenumbers $(K_x,K_z)$ has similar effects.

\subsection{Mean flow and symmetries in PCF and band-free PCF}

\begin{figure}
    \subfloat[PCF]{
    \hspace{4.5em} 
    \includegraphics[width=0.8\columnwidth]{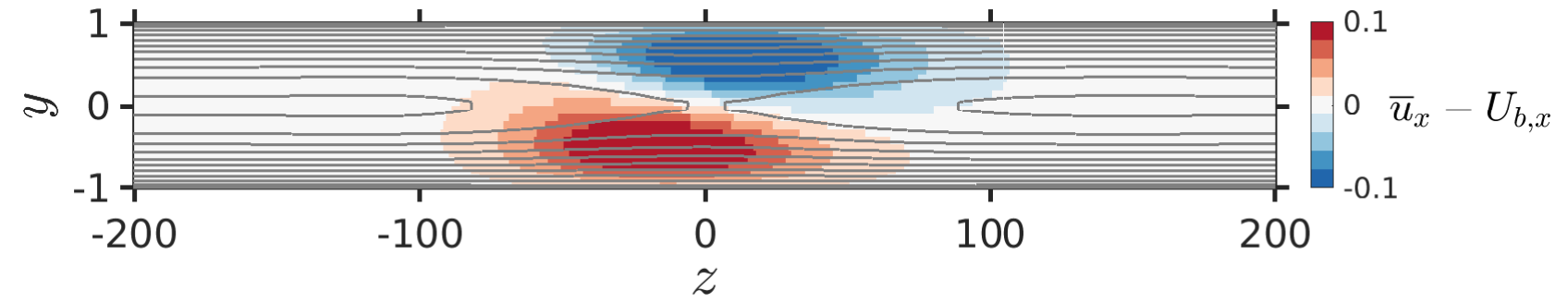}}\\
    \subfloat[Band-free PCF]{\hspace{4.5em} \includegraphics[width=0.8\columnwidth]{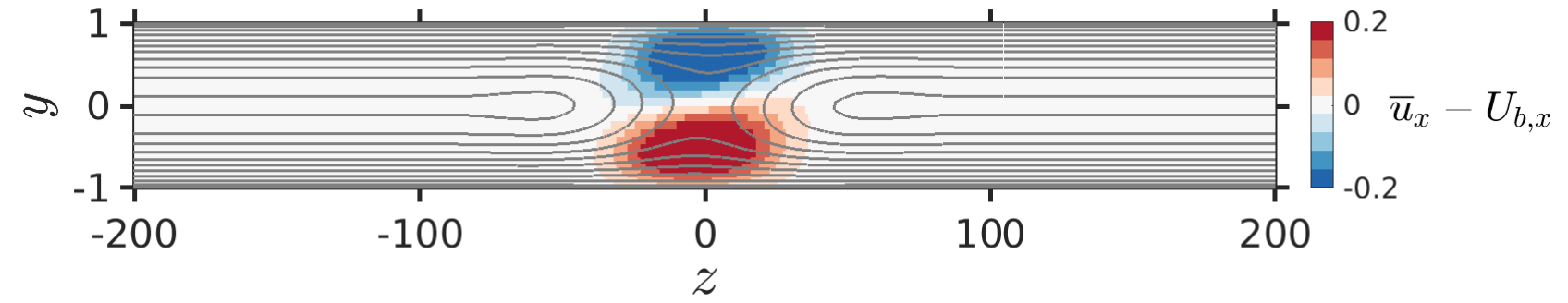}}\\
    \subfloat[PCF]{\hspace{4.5em} \includegraphics[width=0.8\columnwidth]{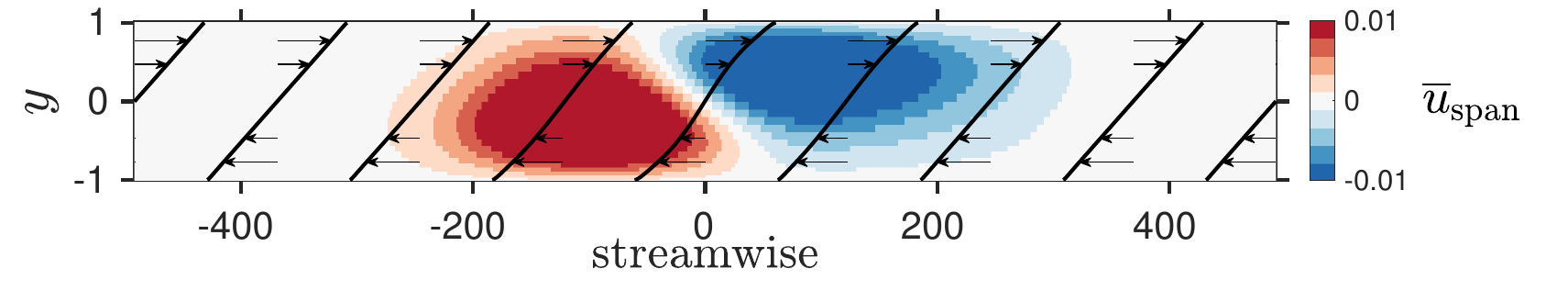}} \\
    \subfloat[Band-free PCF]{\hspace{4.5em} \includegraphics[width=0.8\columnwidth]{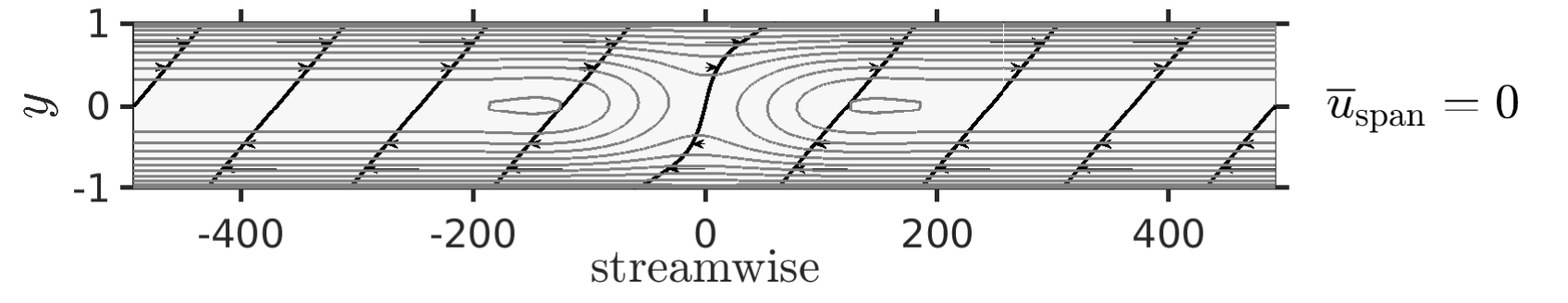}}
    \caption{Mean flow in (a,c) PCF and (b,d) band-free PCF. (a) and (b) show $\overline{u}_x - U_{b,x}$ as colors and  $(\overline{u}_z,\overline{u}_y)$ streamlines in the $(y,z)$ plane. (c) and (d) show 
    $\overline{u}_{\spwise}$ as colors and profiles profiles of $\overline{u}_{\strm}(y)$ in the (streamwise-$y$) plane.
    Additionally, (d) shows streamlines associated to the two-component mean flow $(\overline{u}_\strm,\overline{u}_y)$ for band-free PCF.
    }
    \label{fig:mean_flow}
\end{figure}

In the main paper we discuss the absence of excited and refractory regions for band-free PCF, and how this is related to the absence of discrete structures (bands). Here we provide further details by presenting in Fig.~\ref{fig:mean_flow} the mean flow in PCF (a,c) and band-free PCF (b,d). 
The mean is carried out over time and $x$, for an isolated stationary band in PCF at $Re=340$, and for an approximately stationary localized turbulent patch in band-free PCF at $Re=360$.

Figure~\ref{fig:mean_flow}(a,b) shows the mean flow in the tilted $(z,y)$ plane.
Grey lines are streamlines of $(\overline{u}_z, \overline{u}_y)$ and colors depict $\overline{u}_x - U_{b,x}$, where ${\bf U}_b$ is the laminar flow profile. 
Turbulence has been centered about $z=0$. 
In Fig.~\ref{fig:mean_flow}(c,d), the mean flow is visualized in the non-tilted $(\strm,y)$ plane. Here, colors represent the spanwise component $\overline{u}_{\spwise}$ and turbulence has been centered about $\strm=0$. Profiles $\overline{u}_{\strm} (y)$ are shown as arrows. In PCF, one cannot build a streamfunction in the $(\strm,y)$ plane since the spanwise mean flow varies as a function of the spanwise coordinate.
However, in band-free PCF, $\overline{u}_{\spwise} = 0$, hence it is possible to associate the mean flow in the $(\strm,y)$ plane to a streamfunction. 
The streamlines for band-free PCF, which correspond to the complete mean flow, are shown in Fig.~\ref{fig:mean_flow}(d). 

Note that the wall-normal velocity $\overline{u}_y$ is found to be larger in band-free PCF than in PCF. 
See the $z$-spacing of the streamlines at the interface $z\simeq-20$ in Fig.~\ref{fig:mean_flow}(b), as opposed to that in Fig.~\ref{fig:mean_flow}(a)). 
This can also be seen indirectly comparing the large-scale flow for PCF and band-free PCF in Fig.~1 of the main text. For band-free PCF, the streamwise flow changes rapidly, and in the absence of large-scale spanwise flow, there must be significant wall-normal flow by incompressibility.

Both PCF and band-free PCF exhibit centro-symmetry $(\overline{u}_x, \overline{u}_y, \overline{u}_z)(x,y,z) \to (-\overline{u}_x,-\overline{u}_y,-\overline{u}_z)(-x,-y,-z)$, 
as visible in Figs.~\ref{fig:mean_flow}(a,b).
Centro-symmetry $(\overline{u}_\strm, \overline{u}_y,\overline{u}_\spwise)(\strm,y,\spwise) \to (-\overline{u}_\strm, -\overline{u}_y,-\overline{u}_\spwise)(-\strm,-y,-\spwise)$ is also noticeable in  Figs.~\ref{fig:mean_flow}(c,d). Both centro-symmetry operations are symmetries of the problem, that is symmetries of the equations and boundary conditions.
In band-free PCF, the mean flow appears to present an additional reflection symmetry 
$(\overline{u}_x, \overline{u}_y, \overline{u}_z)(x,y,z) \to (\overline{u}_x,-\overline{u}_y,\overline{u}_z)(x,y,-z)$. 
(Because this is not a symmetry of the PCF configuration, it holds only approximately.)
In particular, the shapes of the color patches corresponding to $\overline{u}_x$ are approximately reflection-symmetric ($z \to -z$) in Fig.~\ref{fig:mean_flow}(b), in contrast to the color patches in Fig.~\ref{fig:mean_flow}(a).
Similarly, in usual streamwise-spanwise coordinates, the symmetry $(\overline{u}_\strm,\overline{u}_y,\overline{u}_\spwise)(\strm,y,\spwise) \to (\overline{u}_\strm,-\overline{u}_y,\overline{u}_\spwise)(-\strm,y,\spwise)$, 
very nearly holds for band-free PCF,
as can be seen in Fig.~\ref{fig:mean_flow}(d).
PCF clearly does not exhibit this approximate symmetry because the mean spanwise flow (color patches in Fig.~\ref{fig:mean_flow}(c)) is not even close to symmetric under streamwise reflection.

The presence of excited and refractory zones, with the associated spatial mean-turbulent feedback (see \citep{barkley2011modeling, hof2010eliminating, gome1, benavides2023model}), is the main reason that shear flows sustain bands or puffs. 
In PCF, the presence of excited and refractory regions is consistent with the absence of reflection symmetry $z\to -z$,  
(or equivalently for pipe flow, with the absence of upstream-downstream symmetry).
Downstream mean flow slowly reacts to upstream turbulence, hence the upstream interface (at $z<0, y>0$ in Fig.~2(a,b) in the main text) differs 
from the downstream interface (at $z>0, y>0$).
Note that the absence of reflection symmetry $z\to -z$ in PCF is related to the \emph{overhang regions} \citep{lundbladh1991direct, duguet2013oblique}, in which laminar flow in one layer faces turbulent flow in the other layer.

In contrast, band-free PCF has a near-reflection symmetry
$z\to -z$ in the mean flow, so that there is no strong difference between upstream and downstream laminar-turbulent interfaces. This implies that there are no excited and refractory regions, and hence no significant overhang regions.
This is why band-free PCF lacks discrete turbulent structures, 
in spite of its having a two-component mean flow like that of pipe flow.
This is true independently of the intensity of $\overline{u}_y$.

Not all wall-bounded shear flows without discrete turbulent structures possess an extra reflection symmetry. 
For example bent pipe flow lacks puffs \citep{rinaldi2019vanishing, zhuang2023discontinuous}, but does not have an approximate upstream-downstream symmetry. 
In such systems, it remains to be seen whether the absence of significant excited and refractory regions is responsible for the absence of discrete turbulent structures.

\captionsetup[subfigure]{format=hang,justification=raggedleft, position=top, font=normalsize, margin= 10pt}

\subsection{Supplement to Figure 4 of the main text: Front contraction speeds}

\begin{figure}[t] \includegraphics[width=0.9\columnwidth]{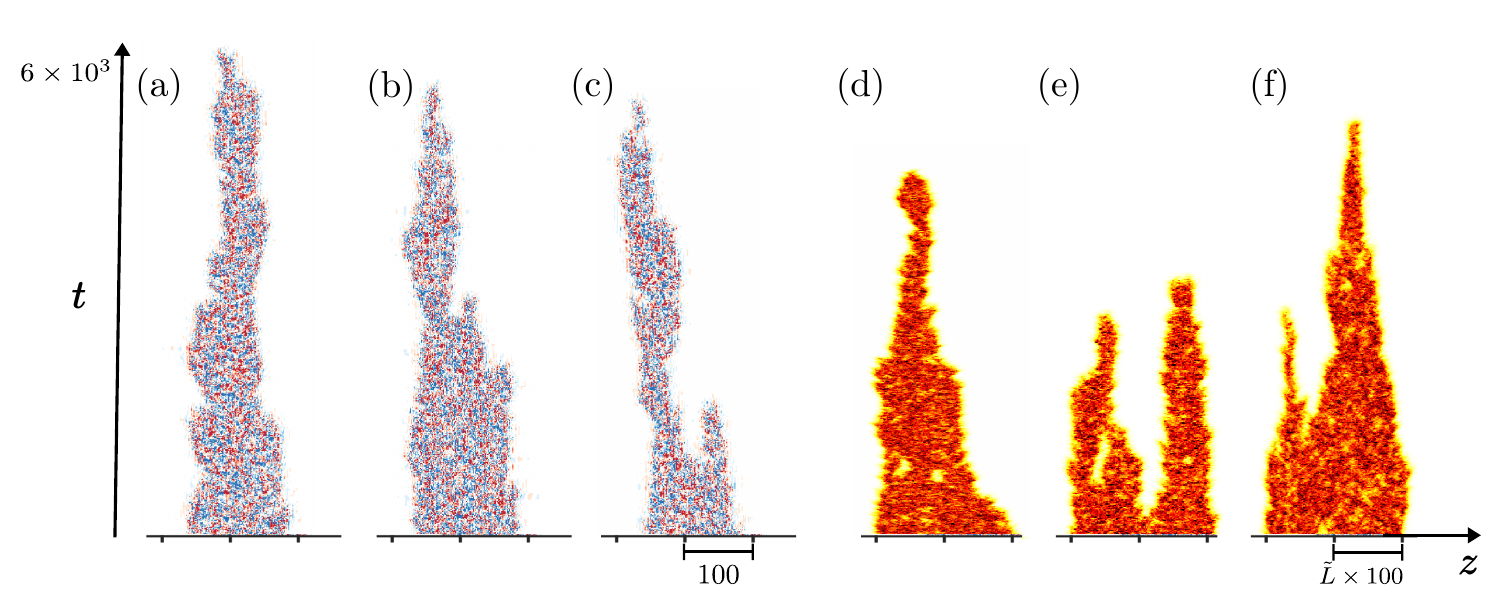} 
\caption{Front contraction below $Re_0$ in band-free PCF at $\theta=24^\circ$ and $Re=381$ (a,b,c), and in the stochastic model considered in the main text (see Eq.~\eqref{eq:model}) with $\sigma=2.0$ (d,e,f), plotted with time and length rescaling and choice of $r$ as discussed in Section~\ref{sec:model_appendix}. Cases shown correspond to different realizations of the same initial condition. In both DNS and model, turbulence contracts and sometimes decays in blocks, making the measurement of a contraction velocity problematic.}
    \label{fig:slug_neg}
\end{figure}

The front velocity measured in Fig.~4 in the main text lacks negative points measuring contraction speeds. The reason is illustrated on Figure \ref{fig:slug_neg}, where a localized turbulent patch is simulated at $Re=381 < Re_0$. The flow laminarizes due to two competing effects: front contraction and the nucleation of laminar gaps 
within the turbulent phase, as seen in Fig.~\ref{fig:slug_neg}c. 
The model in the main text and in Eq.~\eqref{eq:model} below reproduces this phenomenology.
Although a mean negative velocity could be extracted from the mean turbulent lifetime for a given initial width of turbulence, this would not  measure single front contraction, but would average effects of contraction and gap nucleation.

The asymmetry between expanding and contracting turbulent front comes about as follows.
We measure the mean front speed as the change in distance between the most upstream and the most downstream turbulent-laminar fronts, divided by twice the observational time window (to account for the two front motions). 
For sufficiently expanding turbulence, laminar-gap formation within turbulence has no significant effect on this measurement procedure and we can measure mean front speeds precisely. For contracting turbulent patches, even at $Re = 381$, just slightly below $Re_0 = 383$, laminar-gap formation affects our procedure and results in values that are not representative of mean front motion, but rather a combination of front motion and laminar-gap nucleation.

\subsection{Quench experiments}

\begin{figure}
   \includegraphics[width=0.6\columnwidth]{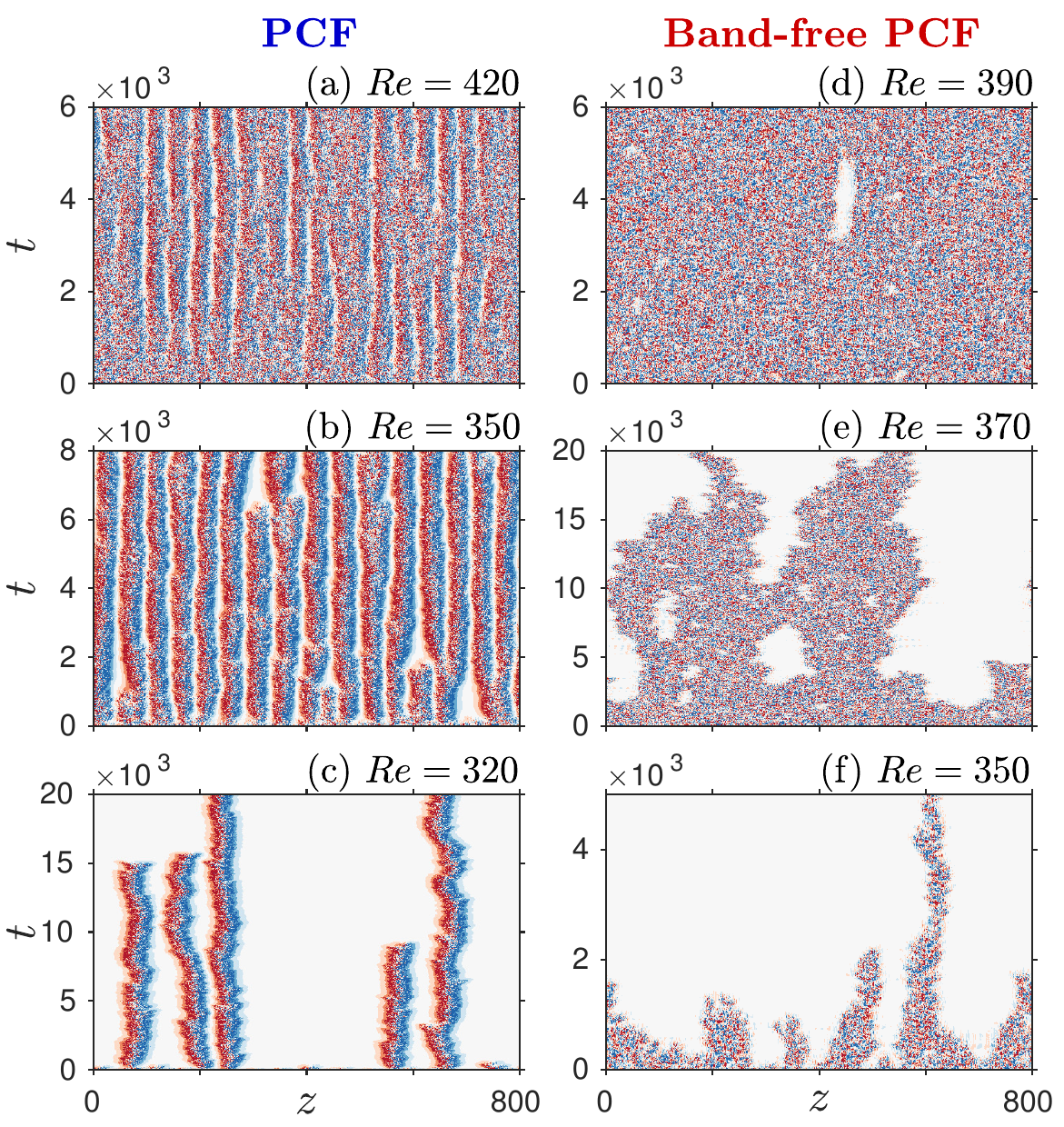} 
   \caption{Quench experiments in both PCF and band-free PCF in a tilted domain with $\theta=24^\circ$. }
   \label{fig:quench}
\end{figure}

We expand on
the quench experiments presented in the main text and show space-time visualizations of the flow at various Reynolds numbers, for both PCF and band-free PCF.
The flow is initiated with uniform turbulence simulated at $Re=500$, and $Re$ is changed to a desired value. PCF spontaneously forms its characteristic isolated bands after a few hundred time units. 
We visualize this process on the left column of Figure \ref{fig:quench} in the tilted domain at angle $\theta=24^\circ$.
Depending on the value of $Re$, the flow resulting from the quench will be made of either intermittent patches of turbulent bands surrounded by an otherwise turbulent flow ($Re=420$), regularly-spaced bands ($Re=350$), or sparse, isolated bands ($Re=320$) which populate the flow 
near the critical point $Re_c\simeq 328.7$ 
\citep{lemoult2016directed}.

In band-free PCF, the flow at $Re=390$ is mostly uniform, although interspersed with 
laminar gaps of various width, which randomly nucleate from the turbulent phase. When $Re$ is decreased, these gaps become wider and contaminate the flow field via fluctuating turbulent-laminar fronts.
This is the equilibrium between front contraction, expansion and gap nucleation that determines the critical $Re$ below which turbulence is not sustained.

\subsection{Stochastic reaction-diffusion model for turbulent front propagation}
\label{sec:model_appendix}

In this section, we elaborate on the stochastic reaction-diffusion model  
and on our fitting procedure with the DNS of band-free PCF.
Let us first consider a dimensional reaction-diffusion model:

\begin{align}
	\tilde{T}\partial_t q & = \tilde{L}^2 \partial^2_{zz} q  -  \partial_q V(q,r) +  (\tilde{L} \tilde{T})^{1/2} \,\sigma \, q \,\xi (z,t)
 \label{eq:model_dim}\\
\intertext{with} V(q,r) &\equiv q^2/2 (1 + (r+1) (q^2/2 - 4q/3))\nonumber
\end{align}
where $\xi(z,t)$ is a space-time white Gaussian noise of unit variance, such that
$<\xi(x,t) \xi (x',t')> = \delta(t-t') \delta(x-x')$.
We nondimensionalize \eqref{eq:model_dim} by the characteristic time $\tilde{T}$ and length $\tilde{L}$, leading to:
\begin{equation}
    \partial_t q  =  \partial_{zz}^2 q - \partial_q V +  \sigma\, q\, \xi
    \label{eq:model}
\end{equation}

\begin{figure}
    \centering
\subfloat[]{ \includegraphics[width=0.5\columnwidth]{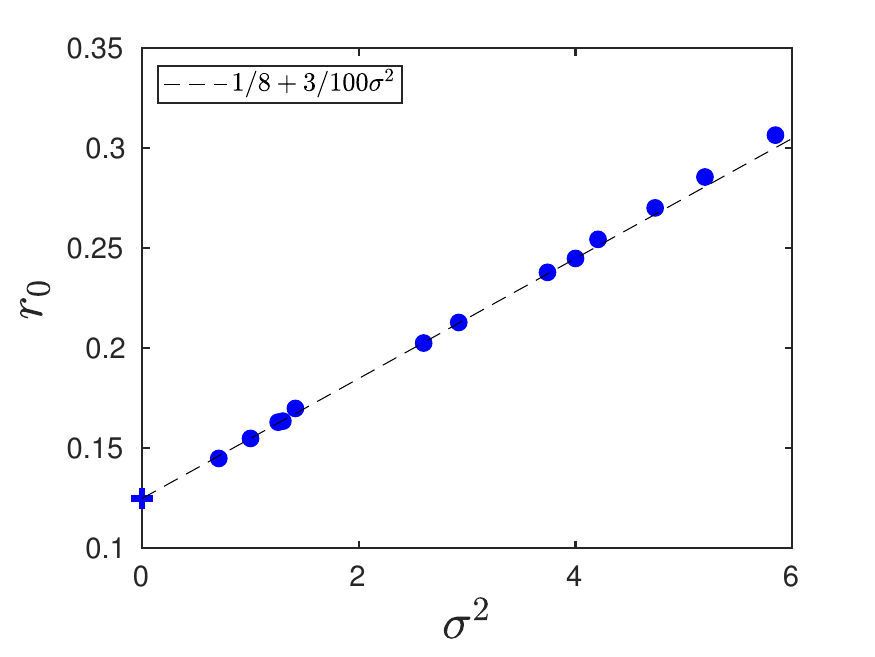} \label{fig:rc}} ~
\subfloat[]{ \includegraphics[width=0.5\columnwidth]{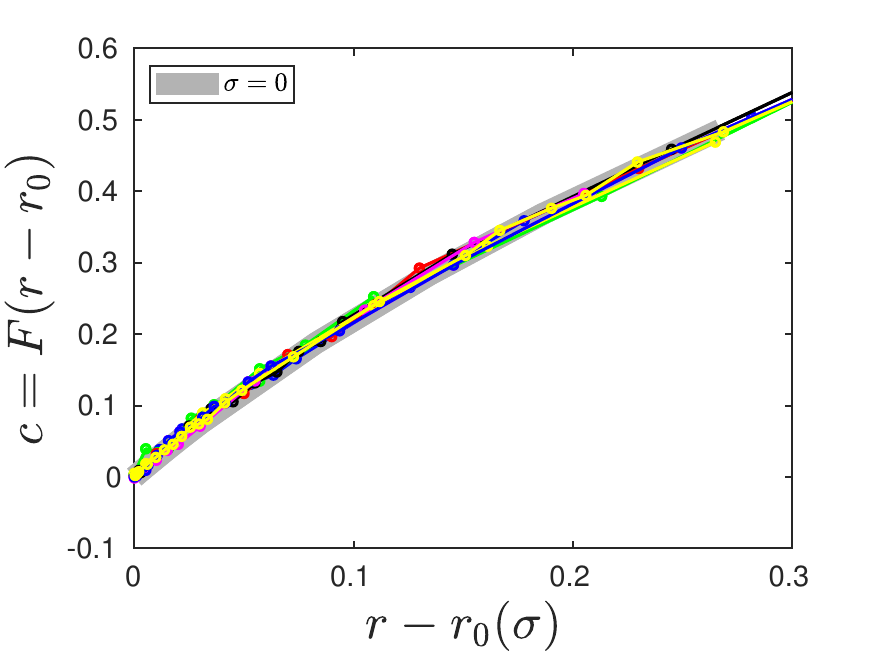} \label{fig:front_model}} 
\caption{(a) Evolution of the zero-propagation point $r_0$ as a function of $\sigma$. The left-most cross shows the deterministic critical point, $r_0=1/8$. Fit $r_0\simeq 1/8 + 3/100 \sigma^2$ is shown as a dashed line.
(b) Front speed $c$ as a function of $r-r_0(\sigma)$ for the values of $\sigma$ shown in (a). All curves collapse to the deterministic function $F$ (thick grey line).}
\end{figure}

The space-independent and deterministic ($\sigma = 0$) version of the model presents a critical value $r= 1/8$ above which the non-zero fixed point is the global minimum of $V$.
When noise is added to the local model, the critical point can be obtained by solving the associated Fokker-Planck equation. Noise both shifts the critical value above which $q\neq0$ is preferred, and also triggers transitions from the turbulent to the absorbing state $q=0$.
When diffusion is added to the problem, 
the solution $q(z,t)$ consists of 
fronts connecting the metastable and the stable fixed points \citep{van2003front}. 
We simulate \eqref{eq:model} with periodic boundary conditions, with a finite-difference scheme under the It\^o representation. 
We measure the expanding front speed from an initial turbulent region and denote $r_0$ the point at which this speed is zero. We find that
\begin{equation}
    r_0 \simeq \frac{1}{8} + \frac{3}{100} \sigma^2
    \label{eq:rc}
\end{equation}
(see Figure \ref{fig:rc}), and that the front speed behaves as (Figure \ref{fig:front_model})
\begin{equation}
c(r, \sigma) \simeq F (r - r_0(\sigma))
\label{eq:c_F}
\end{equation}
with $F$ a non-linear front speed, independent of $\sigma$. $F$ is computed in the deterministic case $\sigma= 0$, via nonlinear simulations (see \citet{barkley2016theoretical}, appendix A).

Note first that \eqref{eq:c_F} is valid for $r>r_0$, because 
the negative front velocities depend on the fluctuations $\sigma$, as discussed in Fig.~\ref{fig:slug_neg}. 
Second, $F$ in Eq.~\eqref{eq:c_F} may depend on $\sigma$ very near $r_0$ depending on the order of the phase transition.
Third, $r_0$ is close to, but {\em a priori} lower than, the critical point $r_c$: when the transition is second-order, the critical point is governed by a balance between slightly expanding fronts and laminar-gap nucleation \citep{hinrichsen2000non}.

From \eqref{eq:rc} and \eqref{eq:c_F}, the dimensional model \eqref{eq:model_dim}, parametrized by $(\tilde{T}, \tilde{L}, r)$, has front speeds obeying the  
rescaled curve
\begin{equation}   c_{\tilde{T},\tilde{L}} (r) = \frac{\tilde{L}}{\tilde{T}} ~ F \left( r - \frac{1}{8} - \frac{3}{100} \sigma^2 \right)
\label{eq:c}
\end{equation}

\begin{figure}
    \centering
    \subfloat[]{ \includegraphics[width=0.35\columnwidth]{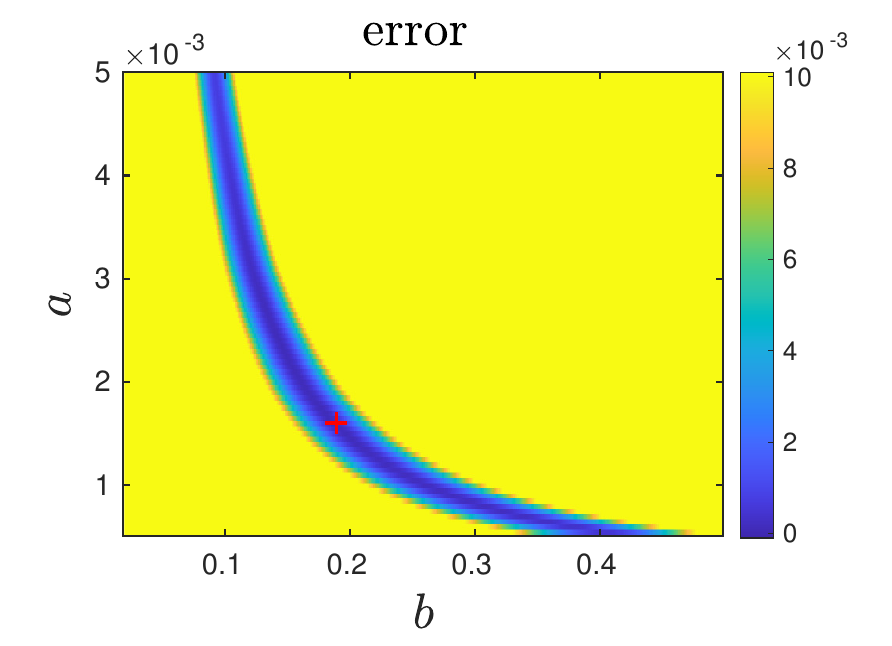} \label{fig:error}}~ 
\subfloat[$\sigma=1$]{ \includegraphics[width=0.33\columnwidth]{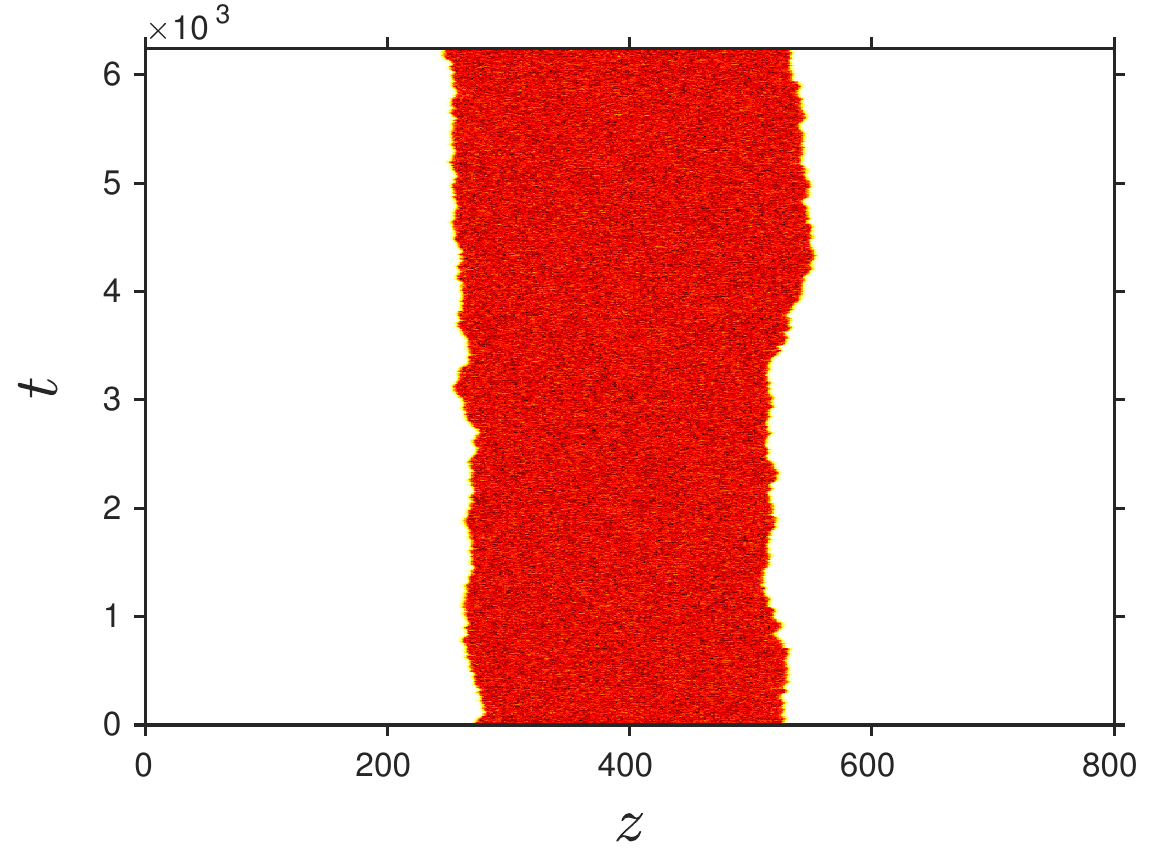}} \label{fig:s1} \\
\subfloat[$\sigma=2.5$]{ \includegraphics[width=0.33\columnwidth]{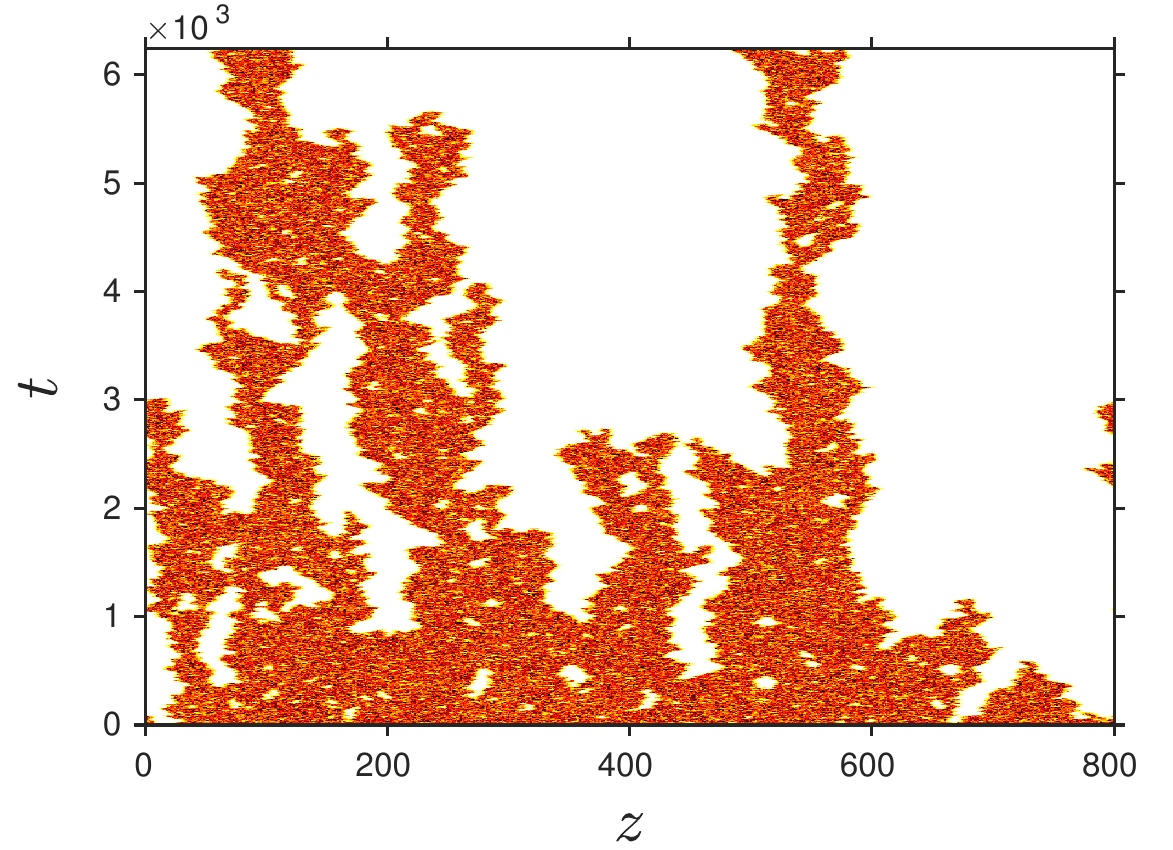} \label{fig:s2.5}} ~
\subfloat[$\sigma=4$]{ \includegraphics[width=0.33\columnwidth]{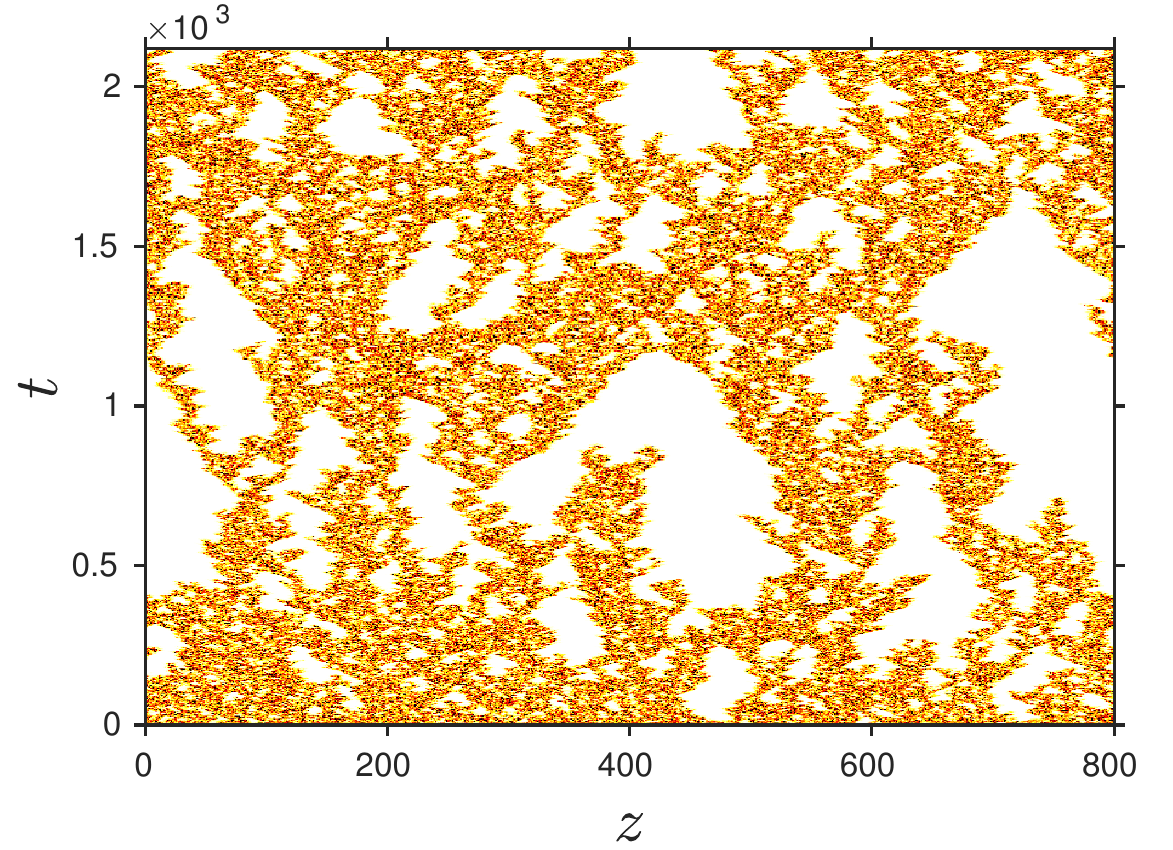} \label{fig:s4}} 
\caption{(a) Mean-square error associated with the fitting  \eqref{eq:fit} with the DNS of band-free PCF. $(a,b)\simeq (0.19,0.0016)$ (red cross) globally minimizes the error. (b,c,d) Visualisation of model \eqref{eq:model} for different values of $\sigma$, with $r$ close to critical point $r_c$. (b) is initiated with a wide localised turbulent zone. As the transition is first order at this value of $\sigma$, the state near critical point is fixed by the initial condition, contrary to (c) and (d).
    }
\end{figure}
We find that this curve is in good agreement with the DNS of band-free PCF, up to a rescaling of the axes $r-r_0$ and $c$ such that
\begin{equation}
c_{\rm DNS} (Re) =  a \: F ( b (Re - Re_0))
\label{eq:fit}
\end{equation}

We first estimate the critical point in the DNS to be $Re_0\simeq 383$ from the zero front speed point.
We then find $a$ and $b$ from fitting the DNS data with the form $F$ extracted from simulations of model \eqref{eq:model}.
The best fit yields $(a, b)=(0.189, 0.0016)$ 
with an mean-squared error of $1.9\times10^{-4}$ (see the solid red line in Fig.~4 in the main text.  Figure \ref{fig:error} shows the mean-squared error associated to this fit. 

From parameter $a$, we find a characteristic velocity scale $\tilde{L} / \tilde{T} = a = c_{\rm DNS}/F$ fitting the DNS front speed.
Similarly, the length scale $\tilde{L}$ can be set to correspond to the front width in PCF. In the DNS, we measure the characteristic front width from Figure 2 in the main text: denoting $\langle E_{\rm turb} \rangle$ the mean turbulent energy in the turbulent zone, we define the front width $l_{\rm DNS}$ such that $E_{\rm turb}$ goes from $0.001 \langle E_{\rm turb} \rangle$ to $0.999 \langle E_{\rm turb} \rangle$. 
We carry out the same measurement in the model (with $q$ instead of $E_{\rm turb}$).
This yields $l_{\rm DNS} \simeq 39.4$ in DNS and $l \simeq 13.74$ in the model (the measurement is carried out with $\sigma=0$).
The ratio of the two gives the rescaling length scale $\tilde{L} = l_{\rm DNS} / l \simeq 3.0$.
The time scale is deduced from $\tilde{T} = \tilde{L} / a \simeq 16$.
Finally, parameter $b$ connects Reynolds-number scales via $r-r_0 = b(Re -Re_0)$ and is used to compare space-time visualizations of model and DNS at the same distance to the zero-speed point.

To find the value of $\sigma$ best corresponding to the DNS, we use the measured turbulent fraction at a finite distance from critical point, 
$Re_c$ in the DNS, and $r_c(\sigma)$ in the model.
Letting $\epsilon_{\rm DNS} \equiv (Re-Re_c)/Re_c$ and $\epsilon \equiv (r-r_c)/r_c$, we choose $\sigma$ such that
\begin{equation}
    F_{t, \rm DNS}  (\epsilon_{\rm DNS} = 0.01) \simeq F_{t, \rm model}  (\epsilon = 0.01, \sigma)
    \label{eq:Ft_0.01}.
\end{equation}
This yields $\sigma=2.0$, and both the DNS and model turbulent fractions are approximately 0.9 at $\epsilon = 0.01$. The precision of this determination is limited by the precision of $Re_c$, which we have only determined to $\pm 0.5$.

\subsection{Supplement to Figure 5 of the main text: turbulent fraction and other DP exponents}

In the DNS of band-free PCF, we define the local turbulent energy as 
\begin{align}
    e(z,t) \equiv \frac{1}{2} \langle u_y(x,y,z,t)^2 + u_\spwise(x,y,z,t)^2 \rangle_{x,y}
\end{align}
and the instantaneous turbulent fraction as 

\begin{align}
   F_t(t) \equiv \int_0^{L_z} dz ~ \Theta(e(z,t) - e_{\rm lam})
\end{align}
where $\Theta$ is the Heaviside function and $e_{\rm lam}$ is a threshold delimiting turbulent and laminar regions. We use $e_{\rm lam}= 3\times 10^{-3}$. The equilibrium turbulent fraction is measured when $F_t(t)$ saturates.
The measurements of $F_t$ in the DNS are done in domains of size $L_z=800$ for $Re$-values far from $Re_c$, and $L_z=2400$ close to $Re_c$. 

In the model, turbulent fraction is defined similarly as
\begin{align}
   F_t(t) \equiv \int_0^L dz ~ \Theta(q(z,t) - q_{\rm lam})
\end{align}
where $q_{\rm lam} = 10^{-3}$. The equilibrium turbulent fraction is measured in a domain of size $L=10^5$.
We find that a different choice of $e_{\rm lam}$ and $q_{\rm lam}$ affects only very weakly the curves presented in Fig.~5(f) in the main text.

\begin{figure}
    \centering
    \subfloat[]{\includegraphics[width=0.5\columnwidth]{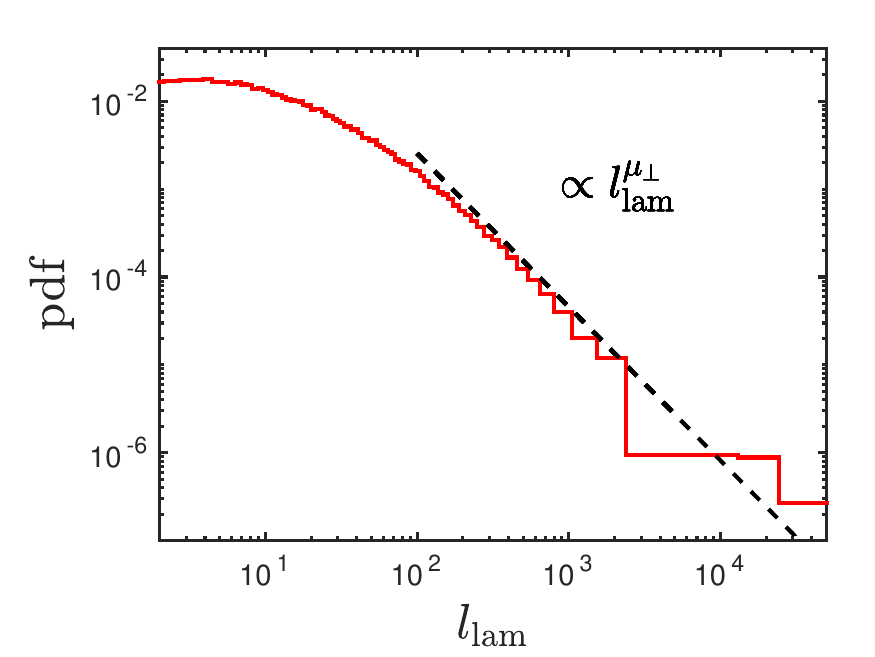}} ~ 
    \subfloat[]{\includegraphics[width=0.5\columnwidth]{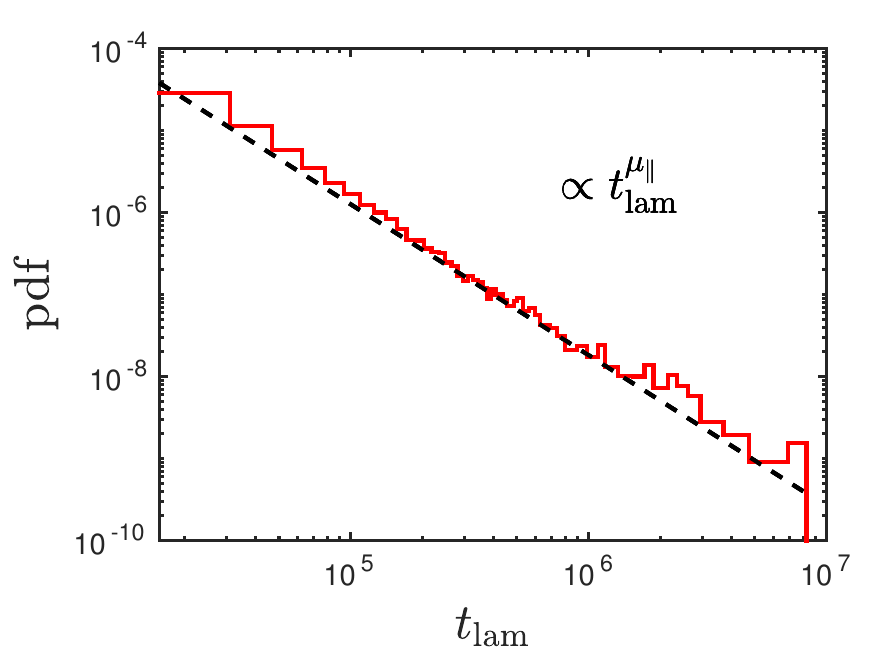}}
    \caption{Scalings for temporal and spatial correlations, from model simulations close to the critical point, $\epsilon = 4\times 10^{-5}$, for noise strength $\sigma=2.0$. Distribution of laminar-gap lengths in (a) the spatial $z$ direction and (b) the temporal direction. Laminar gaps are obtained from simulations in a domain of size $L=10^5$, after the system reaches saturation, for time $t \in [1\times10^6, 2\times 10^7]$.
    Critical power laws for the DP universality class are shown.} 
    \label{fig:DP_expo}
\end{figure}

The DP universality class is characterized by three universal scaling exponents, $\beta$, $\nu_\perp$, $\nu_\parallel$, that dictate the scaling of the 
mean density of active sites and their spatial and temporal correlations in the vicinity of critical point: 
\begin{align}
    F_t \sim \epsilon^\beta, ~~~~ \xi_\perp \sim \epsilon^{-\nu_\perp}, ~~\text{and}   ~~ \xi_\parallel \sim \epsilon^{-\nu_\parallel}
    \label{eq:exponents}
\end{align}
where $\epsilon\equiv (r-r_c)/r_c$ is the reduced control parameter. In one spatial dimension, $\beta\simeq 0.2765$, $\nu_\perp \simeq 1.097$, and $\nu_\parallel \simeq 1.734$.

In Fig.~5(f) of the main text, we have computed the equilibrium turbulent fraction as a function of 
$\epsilon$ for $\sigma=2.0$. The data exhibits scaling consistent with $F_t \sim \epsilon^\beta$.
To assess correlations for this same case, we measure the distributions of empty sites (laminar gaps) in both spatial and temporal directions, giving laminar-gap data $l_{\rm lam}$ and $t_{\rm lam}$. 
The distributions of these data are expected to show power-law behavior $N(l_{\rm lam}) \sim l_{\rm lam}^{\mu_\perp}$ and $N(t_{\rm lam}) \sim t_{\rm lam}^{\mu_\parallel}$, at the critical point. Exponents $\mu_\perp$ and $\mu_\parallel$ are connected to $\nu_\perp$ and $\nu_\parallel$ via
\begin{align}
    \mu_\perp = 2 - \frac{\beta}{\nu_\perp} ~~~ \text{and} ~~~
    \mu_{\parallel} = 2 - \frac{\beta}{\nu_\parallel}.
\end{align}
See \citep{hinrichsen2000non, lubeck2004universal}. In one spatial dimension, $\mu_\perp \simeq 1.748 $ and $\mu_\parallel \simeq 1.840$.

Laminar-gap distributions from model simulations very close to the critical point ($\epsilon = 4\times 10^{-5}$) are shown in Figure~\ref{fig:DP_expo}.
Both distributions exhibit the expected scalings. (The scaling range for the spatial gaps is not large, but a full decade of scaling, $l_{\rm lam} \in [2 \times 10^2, 2 \times 10^3]$, is visible.)
We conclude that for noise amplitude $\sigma = 2.0$, model \eqref{eq:model} exhibits phase transition in the DP universality class. We have not verified the critical exponents associated with correlations at other values of $\sigma$. Nevertheless, in the large-noise cases where we observe a continuous growth in the turbulent fraction with the expected exponent for DP, it is reasonable to assume that these cases are also in the DP universality class, in accordance with the Janssen-Grassberger conjecture \citep{janssen1981nonequilibrium, grassberger1981phase}.

\newpage
\twocolumngrid
\bibliography{bib}

\begin{thebibliography}{60}%
\makeatletter
\providecommand \@ifxundefined [1]{%
 \@ifx{#1\undefined}
}%
\providecommand \@ifnum [1]{%
 \ifnum #1\expandafter \@firstoftwo
 \else \expandafter \@secondoftwo
 \fi
}%
\providecommand \@ifx [1]{%
 \ifx #1\expandafter \@firstoftwo
 \else \expandafter \@secondoftwo
 \fi
}%
\providecommand \natexlab [1]{#1}%
\providecommand \enquote  [1]{``#1''}%
\providecommand \bibnamefont  [1]{#1}%
\providecommand \bibfnamefont [1]{#1}%
\providecommand \citenamefont [1]{#1}%
\providecommand \href@noop [0]{\@secondoftwo}%
\providecommand \href [0]{\begingroup \@sanitize@url \@href}%
\providecommand \@href[1]{\@@startlink{#1}\@@href}%
\providecommand \@@href[1]{\endgroup#1\@@endlink}%
\providecommand \@sanitize@url [0]{\catcode `\\12\catcode `\$12\catcode
  `\&12\catcode `\#12\catcode `\^12\catcode `\_12\catcode `\%12\relax}%
\providecommand \@@startlink[1]{}%
\providecommand \@@endlink[0]{}%
\providecommand \url  [0]{\begingroup\@sanitize@url \@url }%
\providecommand \@url [1]{\endgroup\@href {#1}{\urlprefix }}%
\providecommand \urlprefix  [0]{URL }%
\providecommand \Eprint [0]{\href }%
\providecommand \doibase [0]{http://dx.doi.org/}%
\providecommand \selectlanguage [0]{\@gobble}%
\providecommand \bibinfo  [0]{\@secondoftwo}%
\providecommand \bibfield  [0]{\@secondoftwo}%
\providecommand \translation [1]{[#1]}%
\providecommand \BibitemOpen [0]{}%
\providecommand \bibitemStop [0]{}%
\providecommand \bibitemNoStop [0]{.\EOS\space}%
\providecommand \EOS [0]{\spacefactor3000\relax}%
\providecommand \BibitemShut  [1]{\csname bibitem#1\endcsname}%
\let\auto@bib@innerbib\@empty
\bibitem [{\citenamefont {Landau}\ and\ \citenamefont
  {Lifshitz}(1987)}]{landau1987fluid}%
  \BibitemOpen
  \bibfield  {author} {\bibinfo {author} {\bibfnamefont {L.~D.}\ \bibnamefont
  {Landau}}\ and\ \bibinfo {author} {\bibfnamefont {E.~M.}\ \bibnamefont
  {Lifshitz}},\ }\href@noop {} {\emph {\bibinfo {title} {Fluid Mechanics:
  Course of Theoretical Physics, Volume 6, 2nd ed.}}}\ (\bibinfo  {publisher}
  {Butterworth-Heinemann},\ \bibinfo {year} {1987})\BibitemShut {NoStop}%
\bibitem [{\citenamefont {Pomeau}(1986)}]{pomeau}%
  \BibitemOpen
  \bibfield  {author} {\bibinfo {author} {\bibfnamefont {Y.}~\bibnamefont
  {Pomeau}},\ }\href@noop {} {\bibfield  {journal} {\bibinfo  {journal}
  {Physica D}\ }\textbf {\bibinfo {volume} {23}},\ \bibinfo {pages} {3}
  (\bibinfo {year} {1986})}\BibitemShut {NoStop}%
\bibitem [{\citenamefont {Grassberger}(1981)}]{grassberger1981phase}%
  \BibitemOpen
  \bibfield  {author} {\bibinfo {author} {\bibfnamefont {P.}~\bibnamefont
  {Grassberger}},\ }in\ \href@noop {} {\emph {\bibinfo {booktitle} {Nonlinear
  Phenomena in Chemical Dynamics}}}\ (\bibinfo  {publisher} {Springer},\
  \bibinfo {year} {1981})\ p.\ \bibinfo {pages} {262}\BibitemShut {NoStop}%
\bibitem [{\citenamefont {Janssen}(1981)}]{janssen1981nonequilibrium}%
  \BibitemOpen
  \bibfield  {author} {\bibinfo {author} {\bibfnamefont {H.-K.}\ \bibnamefont
  {Janssen}},\ }\href@noop {} {\bibfield  {journal} {\bibinfo  {journal} {Z.\
  Phys.\ B}\ }\textbf {\bibinfo {volume} {42}},\ \bibinfo {pages} {151}
  (\bibinfo {year} {1981})}\BibitemShut {NoStop}%
\bibitem [{\citenamefont {Waleffe}(1997)}]{waleffe1997self}%
  \BibitemOpen
  \bibfield  {author} {\bibinfo {author} {\bibfnamefont {F.}~\bibnamefont
  {Waleffe}},\ }\href@noop {} {\bibfield  {journal} {\bibinfo  {journal}
  {Phys.\ Fluids}\ }\textbf {\bibinfo {volume} {9}},\ \bibinfo {pages} {883}
  (\bibinfo {year} {1997})}\BibitemShut {NoStop}%
\bibitem [{\citenamefont {Prigent}\ \emph {et~al.}(2002)\citenamefont
  {Prigent}, \citenamefont {Gr{\'e}goire}, \citenamefont {Chat{\'e}},
  \citenamefont {Dauchot},\ and\ \citenamefont {van
  Saarloos}}]{prigent2002large}%
  \BibitemOpen
  \bibfield  {author} {\bibinfo {author} {\bibfnamefont {A.}~\bibnamefont
  {Prigent}}, \bibinfo {author} {\bibfnamefont {G.}~\bibnamefont
  {Gr{\'e}goire}}, \bibinfo {author} {\bibfnamefont {H.}~\bibnamefont
  {Chat{\'e}}}, \bibinfo {author} {\bibfnamefont {O.}~\bibnamefont {Dauchot}},
  \ and\ \bibinfo {author} {\bibfnamefont {W.}~\bibnamefont {van Saarloos}},\
  }\href@noop {} {\bibfield  {journal} {\bibinfo  {journal} {Phys.\ Rev.\
  Lett.}\ }\textbf {\bibinfo {volume} {89}},\ \bibinfo {pages} {014501}
  (\bibinfo {year} {2002})}\BibitemShut {NoStop}%
\bibitem [{\citenamefont {Prigent}\ \emph {et~al.}(2003)\citenamefont
  {Prigent}, \citenamefont {Gr{\'e}goire}, \citenamefont {Chat{\'e}},\ and\
  \citenamefont {Dauchot}}]{prigent2003long}%
  \BibitemOpen
  \bibfield  {author} {\bibinfo {author} {\bibfnamefont {A.}~\bibnamefont
  {Prigent}}, \bibinfo {author} {\bibfnamefont {G.}~\bibnamefont
  {Gr{\'e}goire}}, \bibinfo {author} {\bibfnamefont {H.}~\bibnamefont
  {Chat{\'e}}}, \ and\ \bibinfo {author} {\bibfnamefont {O.}~\bibnamefont
  {Dauchot}},\ }\href@noop {} {\bibfield  {journal} {\bibinfo  {journal}
  {Physica D}\ }\textbf {\bibinfo {volume} {174}},\ \bibinfo {pages} {100}
  (\bibinfo {year} {2003})}\BibitemShut {NoStop}%
\bibitem [{\citenamefont {Barkley}\ and\ \citenamefont
  {Tuckerman}(2005)}]{barkley2005computational}%
  \BibitemOpen
  \bibfield  {author} {\bibinfo {author} {\bibfnamefont {D.}~\bibnamefont
  {Barkley}}\ and\ \bibinfo {author} {\bibfnamefont {L.~S.}\ \bibnamefont
  {Tuckerman}},\ }\href@noop {} {\bibfield  {journal} {\bibinfo  {journal}
  {Phys.\ Rev.\ Lett.}\ }\textbf {\bibinfo {volume} {94}},\ \bibinfo {pages}
  {014502} (\bibinfo {year} {2005})}\BibitemShut {NoStop}%
\bibitem [{\citenamefont {Duguet}\ \emph {et~al.}(2010)\citenamefont {Duguet},
  \citenamefont {Schlatter},\ and\ \citenamefont
  {Henningson}}]{duguet2010formation}%
  \BibitemOpen
  \bibfield  {author} {\bibinfo {author} {\bibfnamefont {Y.}~\bibnamefont
  {Duguet}}, \bibinfo {author} {\bibfnamefont {P.}~\bibnamefont {Schlatter}}, \
  and\ \bibinfo {author} {\bibfnamefont {D.~S.}\ \bibnamefont {Henningson}},\
  }\href@noop {} {\bibfield  {journal} {\bibinfo  {journal} {J.\ Fluid Mech.}\
  }\textbf {\bibinfo {volume} {650}},\ \bibinfo {pages} {119} (\bibinfo {year}
  {2010})}\BibitemShut {NoStop}%
\bibitem [{\citenamefont {Lemoult}\ \emph {et~al.}(2016)\citenamefont
  {Lemoult}, \citenamefont {Shi}, \citenamefont {Avila}, \citenamefont
  {Jalikop}, \citenamefont {Avila},\ and\ \citenamefont
  {Hof}}]{lemoult2016directed}%
  \BibitemOpen
  \bibfield  {author} {\bibinfo {author} {\bibfnamefont {G.}~\bibnamefont
  {Lemoult}}, \bibinfo {author} {\bibfnamefont {L.}~\bibnamefont {Shi}},
  \bibinfo {author} {\bibfnamefont {K.}~\bibnamefont {Avila}}, \bibinfo
  {author} {\bibfnamefont {S.~V.}\ \bibnamefont {Jalikop}}, \bibinfo {author}
  {\bibfnamefont {M.}~\bibnamefont {Avila}}, \ and\ \bibinfo {author}
  {\bibfnamefont {B.}~\bibnamefont {Hof}},\ }\href@noop {} {\bibfield
  {journal} {\bibinfo  {journal} {Nature Physics}\ }\textbf {\bibinfo {volume}
  {12}},\ \bibinfo {pages} {254} (\bibinfo {year} {2016})}\BibitemShut
  {NoStop}%
\bibitem [{\citenamefont {Avila}\ \emph {et~al.}(2011)\citenamefont {Avila},
  \citenamefont {Moxey}, \citenamefont {de~Lozar}, \citenamefont {Avila},
  \citenamefont {Barkley},\ and\ \citenamefont {Hof}}]{avila2011onset}%
  \BibitemOpen
  \bibfield  {author} {\bibinfo {author} {\bibfnamefont {K.}~\bibnamefont
  {Avila}}, \bibinfo {author} {\bibfnamefont {D.}~\bibnamefont {Moxey}},
  \bibinfo {author} {\bibfnamefont {A.}~\bibnamefont {de~Lozar}}, \bibinfo
  {author} {\bibfnamefont {M.}~\bibnamefont {Avila}}, \bibinfo {author}
  {\bibfnamefont {D.}~\bibnamefont {Barkley}}, \ and\ \bibinfo {author}
  {\bibfnamefont {B.}~\bibnamefont {Hof}},\ }\href@noop {} {\bibfield
  {journal} {\bibinfo  {journal} {Science}\ }\textbf {\bibinfo {volume}
  {333}},\ \bibinfo {pages} {192} (\bibinfo {year} {2011})}\BibitemShut
  {NoStop}%
\bibitem [{\citenamefont {Chantry}\ \emph {et~al.}(2017)\citenamefont
  {Chantry}, \citenamefont {Tuckerman},\ and\ \citenamefont
  {Barkley}}]{chantry_universal}%
  \BibitemOpen
  \bibfield  {author} {\bibinfo {author} {\bibfnamefont {M.}~\bibnamefont
  {Chantry}}, \bibinfo {author} {\bibfnamefont {L.~S.}\ \bibnamefont
  {Tuckerman}}, \ and\ \bibinfo {author} {\bibfnamefont {D.}~\bibnamefont
  {Barkley}},\ }\href@noop {} {\bibfield  {journal} {\bibinfo  {journal} {J.\
  Fluid Mech.}\ }\textbf {\bibinfo {volume} {824}},\ \bibinfo {pages} {R1}
  (\bibinfo {year} {2017})}\BibitemShut {NoStop}%
\bibitem [{\citenamefont {Klotz}\ \emph {et~al.}(2022)\citenamefont {Klotz},
  \citenamefont {Lemoult}, \citenamefont {Avila},\ and\ \citenamefont
  {Hof}}]{klotz2022phase}%
  \BibitemOpen
  \bibfield  {author} {\bibinfo {author} {\bibfnamefont {L.}~\bibnamefont
  {Klotz}}, \bibinfo {author} {\bibfnamefont {G.}~\bibnamefont {Lemoult}},
  \bibinfo {author} {\bibfnamefont {K.}~\bibnamefont {Avila}}, \ and\ \bibinfo
  {author} {\bibfnamefont {B.}~\bibnamefont {Hof}},\ }\href@noop {} {\bibfield
  {journal} {\bibinfo  {journal} {Phys.\ Rev.\ Lett.}\ }\textbf {\bibinfo
  {volume} {128}},\ \bibinfo {pages} {014502} (\bibinfo {year}
  {2022})}\BibitemShut {NoStop}%
\bibitem [{\citenamefont {Takeda}\ \emph {et~al.}(2020)\citenamefont {Takeda},
  \citenamefont {Duguet},\ and\ \citenamefont
  {Tsukahara}}]{takeda2020intermittency}%
  \BibitemOpen
  \bibfield  {author} {\bibinfo {author} {\bibfnamefont {K.}~\bibnamefont
  {Takeda}}, \bibinfo {author} {\bibfnamefont {Y.}~\bibnamefont {Duguet}}, \
  and\ \bibinfo {author} {\bibfnamefont {T.}~\bibnamefont {Tsukahara}},\
  }\href@noop {} {\bibfield  {journal} {\bibinfo  {journal} {Entropy}\ }\textbf
  {\bibinfo {volume} {22}},\ \bibinfo {pages} {988} (\bibinfo {year}
  {2020})}\BibitemShut {NoStop}%
\bibitem [{\citenamefont {Kohyama}\ \emph {et~al.}(2022)\citenamefont
  {Kohyama}, \citenamefont {Sano},\ and\ \citenamefont
  {Tsukahara}}]{kohyama2022sidewall}%
  \BibitemOpen
  \bibfield  {author} {\bibinfo {author} {\bibfnamefont {K.}~\bibnamefont
  {Kohyama}}, \bibinfo {author} {\bibfnamefont {M.}~\bibnamefont {Sano}}, \
  and\ \bibinfo {author} {\bibfnamefont {T.}~\bibnamefont {Tsukahara}},\
  }\href@noop {} {\bibfield  {journal} {\bibinfo  {journal} {Phys.\ Fluids}\
  }\textbf {\bibinfo {volume} {34}},\ \bibinfo {pages} {084112} (\bibinfo
  {year} {2022})}\BibitemShut {NoStop}%
\bibitem [{\citenamefont {Coles}\ and\ \citenamefont {van
  Atta}(1966)}]{coles1966progress}%
  \BibitemOpen
  \bibfield  {author} {\bibinfo {author} {\bibfnamefont {D.}~\bibnamefont
  {Coles}}\ and\ \bibinfo {author} {\bibfnamefont {C.}~\bibnamefont {van
  Atta}},\ }\href@noop {} {\bibfield  {journal} {\bibinfo  {journal} {AIAA
  Journal}\ }\textbf {\bibinfo {volume} {4}},\ \bibinfo {pages} {1969}
  (\bibinfo {year} {1966})}\BibitemShut {NoStop}%
\bibitem [{\citenamefont {Wygnanski}\ and\ \citenamefont
  {Champagne}(1973)}]{wygnanski1973transition}%
  \BibitemOpen
  \bibfield  {author} {\bibinfo {author} {\bibfnamefont {I.~J.}\ \bibnamefont
  {Wygnanski}}\ and\ \bibinfo {author} {\bibfnamefont {F.}~\bibnamefont
  {Champagne}},\ }\href@noop {} {\bibfield  {journal} {\bibinfo  {journal} {J.\
  Fluid Mech.}\ }\textbf {\bibinfo {volume} {59}},\ \bibinfo {pages} {281}
  (\bibinfo {year} {1973})}\BibitemShut {NoStop}%
\bibitem [{\citenamefont {Barkley}\ and\ \citenamefont
  {Tuckerman}(2007)}]{barkley2007mean}%
  \BibitemOpen
  \bibfield  {author} {\bibinfo {author} {\bibfnamefont {D.}~\bibnamefont
  {Barkley}}\ and\ \bibinfo {author} {\bibfnamefont {L.~S.}\ \bibnamefont
  {Tuckerman}},\ }\href@noop {} {\bibfield  {journal} {\bibinfo  {journal} {J.\
  Fluid Mech.}\ }\textbf {\bibinfo {volume} {576}},\ \bibinfo {pages} {109}
  (\bibinfo {year} {2007})}\BibitemShut {NoStop}%
\bibitem [{\citenamefont {Duguet}\ and\ \citenamefont
  {Schlatter}(2013)}]{duguet2013oblique}%
  \BibitemOpen
  \bibfield  {author} {\bibinfo {author} {\bibfnamefont {Y.}~\bibnamefont
  {Duguet}}\ and\ \bibinfo {author} {\bibfnamefont {P.}~\bibnamefont
  {Schlatter}},\ }\href@noop {} {\bibfield  {journal} {\bibinfo  {journal}
  {Phys.\ Rev.\ Lett.}\ }\textbf {\bibinfo {volume} {110}},\ \bibinfo {pages}
  {034502} (\bibinfo {year} {2013})}\BibitemShut {NoStop}%
\bibitem [{\citenamefont {Couliou}\ and\ \citenamefont
  {Monchaux}(2015)}]{couliou2015large}%
  \BibitemOpen
  \bibfield  {author} {\bibinfo {author} {\bibfnamefont {M.}~\bibnamefont
  {Couliou}}\ and\ \bibinfo {author} {\bibfnamefont {R.}~\bibnamefont
  {Monchaux}},\ }\href@noop {} {\bibfield  {journal} {\bibinfo  {journal}
  {Phys.\ Fluids}\ }\textbf {\bibinfo {volume} {27}},\ \bibinfo {pages}
  {034101} (\bibinfo {year} {2015})}\BibitemShut {NoStop}%
\bibitem [{\citenamefont {Klotz}\ \emph {et~al.}(2021)\citenamefont {Klotz},
  \citenamefont {Pavlenko},\ and\ \citenamefont
  {Wesfreid}}]{klotz2021experimental}%
  \BibitemOpen
  \bibfield  {author} {\bibinfo {author} {\bibfnamefont {L.}~\bibnamefont
  {Klotz}}, \bibinfo {author} {\bibfnamefont {A.}~\bibnamefont {Pavlenko}}, \
  and\ \bibinfo {author} {\bibfnamefont {J.}~\bibnamefont {Wesfreid}},\
  }\href@noop {} {\bibfield  {journal} {\bibinfo  {journal} {J.\ Fluid Mech.}\
  }\textbf {\bibinfo {volume} {912}} (\bibinfo {year} {2021})}\BibitemShut
  {NoStop}%
\bibitem [{\citenamefont {Marensi}\ \emph {et~al.}(2023)\citenamefont
  {Marensi}, \citenamefont {Yaln{\i}z},\ and\ \citenamefont
  {Hof}}]{marensi2023dynamics}%
  \BibitemOpen
  \bibfield  {author} {\bibinfo {author} {\bibfnamefont {E.}~\bibnamefont
  {Marensi}}, \bibinfo {author} {\bibfnamefont {G.}~\bibnamefont {Yaln{\i}z}},
  \ and\ \bibinfo {author} {\bibfnamefont {B.}~\bibnamefont {Hof}},\
  }\href@noop {} {\bibfield  {journal} {\bibinfo  {journal} {J.\ Fluid Mech.}\
  }\textbf {\bibinfo {volume} {974}},\ \bibinfo {pages} {A21} (\bibinfo {year}
  {2023})}\BibitemShut {NoStop}%
\bibitem [{\citenamefont {van Doorne}\ and\ \citenamefont
  {Westerweel}(2009)}]{van2009flow}%
  \BibitemOpen
  \bibfield  {author} {\bibinfo {author} {\bibfnamefont {C.~W.}\ \bibnamefont
  {van Doorne}}\ and\ \bibinfo {author} {\bibfnamefont {J.}~\bibnamefont
  {Westerweel}},\ }\href@noop {} {\bibfield  {journal} {\bibinfo  {journal}
  {Philos.\ Trans.\ R.\ Soc.\ A}\ }\textbf {\bibinfo {volume} {367}},\ \bibinfo
  {pages} {489} (\bibinfo {year} {2009})}\BibitemShut {NoStop}%
\bibitem [{\citenamefont {Hof}\ \emph {et~al.}(2010)\citenamefont {Hof},
  \citenamefont {De~Lozar}, \citenamefont {Avila}, \citenamefont {Tu},\ and\
  \citenamefont {Schneider}}]{hof2010eliminating}%
  \BibitemOpen
  \bibfield  {author} {\bibinfo {author} {\bibfnamefont {B.}~\bibnamefont
  {Hof}}, \bibinfo {author} {\bibfnamefont {A.}~\bibnamefont {De~Lozar}},
  \bibinfo {author} {\bibfnamefont {M.}~\bibnamefont {Avila}}, \bibinfo
  {author} {\bibfnamefont {X.}~\bibnamefont {Tu}}, \ and\ \bibinfo {author}
  {\bibfnamefont {T.~M.}\ \bibnamefont {Schneider}},\ }\href@noop {} {\bibfield
   {journal} {\bibinfo  {journal} {Science}\ }\textbf {\bibinfo {volume}
  {327}},\ \bibinfo {pages} {1491} (\bibinfo {year} {2010})}\BibitemShut
  {NoStop}%
\bibitem [{\citenamefont {Samanta}\ \emph {et~al.}(2011)\citenamefont
  {Samanta}, \citenamefont {De~Lozar},\ and\ \citenamefont
  {Hof}}]{samanta2011experimental}%
  \BibitemOpen
  \bibfield  {author} {\bibinfo {author} {\bibfnamefont {D.}~\bibnamefont
  {Samanta}}, \bibinfo {author} {\bibfnamefont {A.}~\bibnamefont {De~Lozar}}, \
  and\ \bibinfo {author} {\bibfnamefont {B.}~\bibnamefont {Hof}},\ }\href@noop
  {} {\bibfield  {journal} {\bibinfo  {journal} {J.\ Fluid Mech.}\ }\textbf
  {\bibinfo {volume} {681}},\ \bibinfo {pages} {193} (\bibinfo {year}
  {2011})}\BibitemShut {NoStop}%
\bibitem [{\citenamefont {Song}\ \emph {et~al.}(2017)\citenamefont {Song},
  \citenamefont {Barkley}, \citenamefont {Hof},\ and\ \citenamefont
  {Avila}}]{song2017speed}%
  \BibitemOpen
  \bibfield  {author} {\bibinfo {author} {\bibfnamefont {B.}~\bibnamefont
  {Song}}, \bibinfo {author} {\bibfnamefont {D.}~\bibnamefont {Barkley}},
  \bibinfo {author} {\bibfnamefont {B.}~\bibnamefont {Hof}}, \ and\ \bibinfo
  {author} {\bibfnamefont {M.}~\bibnamefont {Avila}},\ }\href@noop {}
  {\bibfield  {journal} {\bibinfo  {journal} {J.\ Fluid Mech.}\ }\textbf
  {\bibinfo {volume} {813}},\ \bibinfo {pages} {1045} (\bibinfo {year}
  {2017})}\BibitemShut {NoStop}%
\bibitem [{\citenamefont {Gom{\'e}}\ \emph
  {et~al.}(2023{\natexlab{a}})\citenamefont {Gom{\'e}}, \citenamefont
  {Tuckerman},\ and\ \citenamefont {Barkley}}]{gome2}%
  \BibitemOpen
  \bibfield  {author} {\bibinfo {author} {\bibfnamefont {S.}~\bibnamefont
  {Gom{\'e}}}, \bibinfo {author} {\bibfnamefont {L.~S.}\ \bibnamefont
  {Tuckerman}}, \ and\ \bibinfo {author} {\bibfnamefont {D.}~\bibnamefont
  {Barkley}},\ }\href@noop {} {\bibfield  {journal} {\bibinfo  {journal} {J.\
  Fluid Mech.}\ }\textbf {\bibinfo {volume} {964}},\ \bibinfo {pages} {A17}
  (\bibinfo {year} {2023}{\natexlab{a}})}\BibitemShut {NoStop}%
\bibitem [{\citenamefont {Gom{\'e}}\ \emph
  {et~al.}(2023{\natexlab{b}})\citenamefont {Gom{\'e}}, \citenamefont
  {Tuckerman},\ and\ \citenamefont {Barkley}}]{gome1}%
  \BibitemOpen
  \bibfield  {author} {\bibinfo {author} {\bibfnamefont {S.}~\bibnamefont
  {Gom{\'e}}}, \bibinfo {author} {\bibfnamefont {L.~S.}\ \bibnamefont
  {Tuckerman}}, \ and\ \bibinfo {author} {\bibfnamefont {D.}~\bibnamefont
  {Barkley}},\ }\href@noop {} {\bibfield  {journal} {\bibinfo  {journal} {J.\
  Fluid Mech.}\ }\textbf {\bibinfo {volume} {964}},\ \bibinfo {pages} {A16}
  (\bibinfo {year} {2023}{\natexlab{b}})}\BibitemShut {NoStop}%
\bibitem [{\citenamefont
  {Barkley}(2011{\natexlab{a}})}]{barkley2011simplifying}%
  \BibitemOpen
  \bibfield  {author} {\bibinfo {author} {\bibfnamefont {D.}~\bibnamefont
  {Barkley}},\ }\href@noop {} {\bibfield  {journal} {\bibinfo  {journal}
  {Phys.\ Rev.\ E}\ }\textbf {\bibinfo {volume} {84}},\ \bibinfo {pages}
  {016309} (\bibinfo {year} {2011}{\natexlab{a}})}\BibitemShut {NoStop}%
\bibitem [{\citenamefont {Barkley}(2016)}]{barkley2016theoretical}%
  \BibitemOpen
  \bibfield  {author} {\bibinfo {author} {\bibfnamefont {D.}~\bibnamefont
  {Barkley}},\ }\href@noop {} {\bibfield  {journal} {\bibinfo  {journal} {J.\
  Fluid Mech.}\ }\textbf {\bibinfo {volume} {803}},\ \bibinfo {pages} {P1}
  (\bibinfo {year} {2016})}\BibitemShut {NoStop}%
\bibitem [{\citenamefont {Benavides}\ and\ \citenamefont
  {Barkley}(2023)}]{benavides2023model}%
  \BibitemOpen
  \bibfield  {author} {\bibinfo {author} {\bibfnamefont {S.~J.}\ \bibnamefont
  {Benavides}}\ and\ \bibinfo {author} {\bibfnamefont {D.}~\bibnamefont
  {Barkley}},\ }\href@noop {} {\bibfield  {journal} {\bibinfo  {journal}
  {arXiv:2309.12879}\ } (\bibinfo {year} {2023})}\BibitemShut {NoStop}%
\bibitem [{\citenamefont {Barkley}(2011{\natexlab{b}})}]{barkley2011modeling}%
  \BibitemOpen
  \bibfield  {author} {\bibinfo {author} {\bibfnamefont {D.}~\bibnamefont
  {Barkley}},\ }\href@noop {} {\bibfield  {journal} {\bibinfo  {journal} {J.\
  Phys.: Conf.\ Ser.}\ }\textbf {\bibinfo {volume} {318}},\ \bibinfo {pages}
  {032001} (\bibinfo {year} {2011}{\natexlab{b}})}\BibitemShut {NoStop}%
\bibitem [{\citenamefont {Wang}\ \emph {et~al.}(2022)\citenamefont {Wang},
  \citenamefont {Shih},\ and\ \citenamefont {Goldenfeld}}]{wang2022stochastic}%
  \BibitemOpen
  \bibfield  {author} {\bibinfo {author} {\bibfnamefont {X.}~\bibnamefont
  {Wang}}, \bibinfo {author} {\bibfnamefont {H.-Y.}\ \bibnamefont {Shih}}, \
  and\ \bibinfo {author} {\bibfnamefont {N.}~\bibnamefont {Goldenfeld}},\
  }\href@noop {} {\bibfield  {journal} {\bibinfo  {journal} {Phys.\ Rev.\
  Lett.}\ }\textbf {\bibinfo {volume} {129}},\ \bibinfo {pages} {034501}
  (\bibinfo {year} {2022})}\BibitemShut {NoStop}%
\bibitem [{\citenamefont {Gibson}\ \emph {et~al.}()\citenamefont {Gibson},
  \citenamefont {Reetz}, \citenamefont {Azimi}, \citenamefont {Ferraro},
  \citenamefont {Kreilos}, \citenamefont {Schrobsdorff}, \citenamefont
  {Farano}, \citenamefont {Yesil}, \citenamefont {Sch\"utz}, \citenamefont
  {Culpo},\ and\ \citenamefont {Schneider}}]{channelflow}%
  \BibitemOpen
  \bibfield  {author} {\bibinfo {author} {\bibfnamefont {J.}~\bibnamefont
  {Gibson}}, \bibinfo {author} {\bibfnamefont {F.}~\bibnamefont {Reetz}},
  \bibinfo {author} {\bibfnamefont {S.}~\bibnamefont {Azimi}}, \bibinfo
  {author} {\bibfnamefont {A.}~\bibnamefont {Ferraro}}, \bibinfo {author}
  {\bibfnamefont {T.}~\bibnamefont {Kreilos}}, \bibinfo {author} {\bibfnamefont
  {H.}~\bibnamefont {Schrobsdorff}}, \bibinfo {author} {\bibfnamefont
  {M.}~\bibnamefont {Farano}}, \bibinfo {author} {\bibfnamefont
  {A.}~\bibnamefont {Yesil}}, \bibinfo {author} {\bibfnamefont
  {S.}~\bibnamefont {Sch\"utz}}, \bibinfo {author} {\bibfnamefont
  {M.}~\bibnamefont {Culpo}}, \ and\ \bibinfo {author} {\bibfnamefont
  {T.}~\bibnamefont {Schneider}},\ }\href@noop {} {\enquote {\bibinfo {title}
  {Channelflow 2.0},}\ }\bibinfo {note} {Manuscript in preparation, see
  channelflow.ch}\BibitemShut {NoStop}%
\bibitem [{\citenamefont {Hamilton}\ \emph {et~al.}(1995)\citenamefont
  {Hamilton}, \citenamefont {Kim},\ and\ \citenamefont
  {Waleffe}}]{hamilton1995regeneration}%
  \BibitemOpen
  \bibfield  {author} {\bibinfo {author} {\bibfnamefont {J.~M.}\ \bibnamefont
  {Hamilton}}, \bibinfo {author} {\bibfnamefont {J.}~\bibnamefont {Kim}}, \
  and\ \bibinfo {author} {\bibfnamefont {F.}~\bibnamefont {Waleffe}},\
  }\href@noop {} {\bibfield  {journal} {\bibinfo  {journal} {J.\ Fluid Mech.}\
  }\textbf {\bibinfo {volume} {287}},\ \bibinfo {pages} {317} (\bibinfo {year}
  {1995})}\BibitemShut {NoStop}%
\bibitem [{\citenamefont {Liu}\ \emph {et~al.}(2024)\citenamefont {Liu},
  \citenamefont {Semin}, \citenamefont {Godoy-Diana},\ and\ \citenamefont
  {Wesfreid}}]{liu2024lift}%
  \BibitemOpen
  \bibfield  {author} {\bibinfo {author} {\bibfnamefont {T.}~\bibnamefont
  {Liu}}, \bibinfo {author} {\bibfnamefont {B.}~\bibnamefont {Semin}}, \bibinfo
  {author} {\bibfnamefont {R.}~\bibnamefont {Godoy-Diana}}, \ and\ \bibinfo
  {author} {\bibfnamefont {J.~E.}\ \bibnamefont {Wesfreid}},\ }\href@noop {}
  {\bibfield  {journal} {\bibinfo  {journal} {Phys.\ Rev.\ Fluids}\ }\textbf
  {\bibinfo {volume} {9}},\ \bibinfo {pages} {033901} (\bibinfo {year}
  {2024})}\BibitemShut {NoStop}%
\bibitem [{gom()}]{gomeSMarxiv}%
  \BibitemOpen
  \href@noop {} {}\bibinfo {note} {Appendix contains additional information on
  numerical methods and model \eqref{eq:DP_model} and Figs.\ \ref{fig:mf},
  \ref{fig:slug}, and \ref{fig:model_vs_dns}.}\BibitemShut {Stop}%
\bibitem [{\citenamefont {Shi}\ \emph {et~al.}(2013)\citenamefont {Shi},
  \citenamefont {Avila},\ and\ \citenamefont {Hof}}]{shi}%
  \BibitemOpen
  \bibfield  {author} {\bibinfo {author} {\bibfnamefont {L.}~\bibnamefont
  {Shi}}, \bibinfo {author} {\bibfnamefont {M.}~\bibnamefont {Avila}}, \ and\
  \bibinfo {author} {\bibfnamefont {B.}~\bibnamefont {Hof}},\ }\href@noop {}
  {\bibfield  {journal} {\bibinfo  {journal} {Phys.\ Rev.\ Lett.}\ }\textbf
  {\bibinfo {volume} {110}},\ \bibinfo {pages} {204502} (\bibinfo {year}
  {2013})}\BibitemShut {NoStop}%
\bibitem [{\citenamefont {Gom{\'e}}\ \emph {et~al.}(2020)\citenamefont
  {Gom{\'e}}, \citenamefont {Tuckerman},\ and\ \citenamefont
  {Barkley}}]{gome2020statistical}%
  \BibitemOpen
  \bibfield  {author} {\bibinfo {author} {\bibfnamefont {S.}~\bibnamefont
  {Gom{\'e}}}, \bibinfo {author} {\bibfnamefont {L.~S.}\ \bibnamefont
  {Tuckerman}}, \ and\ \bibinfo {author} {\bibfnamefont {D.}~\bibnamefont
  {Barkley}},\ }\href@noop {} {\bibfield  {journal} {\bibinfo  {journal}
  {Phys.\ Rev.\ Fluids}\ }\textbf {\bibinfo {volume} {5}},\ \bibinfo {pages}
  {083905} (\bibinfo {year} {2020})}\BibitemShut {NoStop}%
\bibitem [{\citenamefont {Avila}\ \emph {et~al.}(2010)\citenamefont {Avila},
  \citenamefont {Willis},\ and\ \citenamefont {Hof}}]{avila2010transient}%
  \BibitemOpen
  \bibfield  {author} {\bibinfo {author} {\bibfnamefont {M.}~\bibnamefont
  {Avila}}, \bibinfo {author} {\bibfnamefont {A.~P.}\ \bibnamefont {Willis}}, \
  and\ \bibinfo {author} {\bibfnamefont {B.}~\bibnamefont {Hof}},\ }\href@noop
  {} {\bibfield  {journal} {\bibinfo  {journal} {J.\ Fluid Mech.}\ }\textbf
  {\bibinfo {volume} {646}},\ \bibinfo {pages} {127} (\bibinfo {year}
  {2010})}\BibitemShut {NoStop}%
\bibitem [{\citenamefont {Mukund}\ \emph {et~al.}(2021)\citenamefont {Mukund},
  \citenamefont {Paranjape}, \citenamefont {Sitte},\ and\ \citenamefont
  {Hof}}]{mukund2021aging}%
  \BibitemOpen
  \bibfield  {author} {\bibinfo {author} {\bibfnamefont {V.}~\bibnamefont
  {Mukund}}, \bibinfo {author} {\bibfnamefont {C.}~\bibnamefont {Paranjape}},
  \bibinfo {author} {\bibfnamefont {M.~P.}\ \bibnamefont {Sitte}}, \ and\
  \bibinfo {author} {\bibfnamefont {B.}~\bibnamefont {Hof}},\ }\href@noop {}
  {\bibfield  {journal} {\bibinfo  {journal} {arXiv:2112.06537}\ } (\bibinfo
  {year} {2021})}\BibitemShut {NoStop}%
\bibitem [{\citenamefont {Xu}\ and\ \citenamefont {Song}(2022)}]{xu2022size}%
  \BibitemOpen
  \bibfield  {author} {\bibinfo {author} {\bibfnamefont {D.}~\bibnamefont
  {Xu}}\ and\ \bibinfo {author} {\bibfnamefont {B.}~\bibnamefont {Song}},\
  }\href@noop {} {\bibfield  {journal} {\bibinfo  {journal} {J.\ Fluid Mech.}\
  }\textbf {\bibinfo {volume} {950}},\ \bibinfo {pages} {R3} (\bibinfo {year}
  {2022})}\BibitemShut {NoStop}%
\bibitem [{\citenamefont {Pomeau}(2015)}]{pomeau2015transition}%
  \BibitemOpen
  \bibfield  {author} {\bibinfo {author} {\bibfnamefont {Y.}~\bibnamefont
  {Pomeau}},\ }\href@noop {} {\bibfield  {journal} {\bibinfo  {journal}
  {Comptes Rendus M{\'e}canique}\ }\textbf {\bibinfo {volume} {343}},\ \bibinfo
  {pages} {210} (\bibinfo {year} {2015})}\BibitemShut {NoStop}%
\bibitem [{\citenamefont {Munoz}(1998)}]{munoz1998nature}%
  \BibitemOpen
  \bibfield  {author} {\bibinfo {author} {\bibfnamefont {M.~A.}\ \bibnamefont
  {Munoz}},\ }\href@noop {} {\bibfield  {journal} {\bibinfo  {journal} {Phys.\
  Rev.\ E}\ }\textbf {\bibinfo {volume} {57}},\ \bibinfo {pages} {1377}
  (\bibinfo {year} {1998})}\BibitemShut {NoStop}%
\bibitem [{\citenamefont {Munoz}\ and\ \citenamefont
  {Pastor-Satorras}(2003)}]{munoz2003stochastic}%
  \BibitemOpen
  \bibfield  {author} {\bibinfo {author} {\bibfnamefont {M.~A.}\ \bibnamefont
  {Munoz}}\ and\ \bibinfo {author} {\bibfnamefont {R.}~\bibnamefont
  {Pastor-Satorras}},\ }\href@noop {} {\bibfield  {journal} {\bibinfo
  {journal} {Phys.\ Rev.\ Lett.}\ }\textbf {\bibinfo {volume} {90}},\ \bibinfo
  {pages} {204101} (\bibinfo {year} {2003})}\BibitemShut {NoStop}%
\bibitem [{\citenamefont {Munoz}(2003)}]{munoz2003multiplicative}%
  \BibitemOpen
  \bibfield  {author} {\bibinfo {author} {\bibfnamefont {M.~A.}\ \bibnamefont
  {Munoz}},\ }\href@noop {} {\bibfield  {journal} {\bibinfo  {journal}
  {arXiv:cond-mat/0303650}\ } (\bibinfo {year} {2003})}\BibitemShut {NoStop}%
\bibitem [{\citenamefont {Gardiner}(1985)}]{gardiner1985handbook}%
  \BibitemOpen
  \bibfield  {author} {\bibinfo {author} {\bibfnamefont {C.~W.}\ \bibnamefont
  {Gardiner}},\ }\href@noop {} {\emph {\bibinfo {title} {Handbook of stochastic
  methods}}},\ Vol.~\bibinfo {volume} {3}\ (\bibinfo  {publisher} {Springer
  Berlin},\ \bibinfo {year} {1985})\BibitemShut {NoStop}%
\bibitem [{\citenamefont {Rinaldi}\ \emph {et~al.}(2019)\citenamefont
  {Rinaldi}, \citenamefont {Canton},\ and\ \citenamefont
  {Schlatter}}]{rinaldi2019vanishing}%
  \BibitemOpen
  \bibfield  {author} {\bibinfo {author} {\bibfnamefont {E.}~\bibnamefont
  {Rinaldi}}, \bibinfo {author} {\bibfnamefont {J.}~\bibnamefont {Canton}}, \
  and\ \bibinfo {author} {\bibfnamefont {P.}~\bibnamefont {Schlatter}},\
  }\href@noop {} {\bibfield  {journal} {\bibinfo  {journal} {J.\ Fluid Mech.}\
  }\textbf {\bibinfo {volume} {866}},\ \bibinfo {pages} {487} (\bibinfo {year}
  {2019})}\BibitemShut {NoStop}%
\bibitem [{\citenamefont {Zhuang}\ \emph {et~al.}(2023)\citenamefont {Zhuang},
  \citenamefont {Yang}, \citenamefont {Mukund}, \citenamefont {Marensi},\ and\
  \citenamefont {Hof}}]{zhuang2023discontinuous}%
  \BibitemOpen
  \bibfield  {author} {\bibinfo {author} {\bibfnamefont {Y.}~\bibnamefont
  {Zhuang}}, \bibinfo {author} {\bibfnamefont {B.}~\bibnamefont {Yang}},
  \bibinfo {author} {\bibfnamefont {V.}~\bibnamefont {Mukund}}, \bibinfo
  {author} {\bibfnamefont {E.}~\bibnamefont {Marensi}}, \ and\ \bibinfo
  {author} {\bibfnamefont {B.}~\bibnamefont {Hof}},\ }\href@noop {} {\bibfield
  {journal} {\bibinfo  {journal} {arXiv:2311.11474}\ } (\bibinfo {year}
  {2023})}\BibitemShut {NoStop}%
\bibitem [{\citenamefont {Rorai}\ \emph {et~al.}(2014)\citenamefont {Rorai},
  \citenamefont {Mininni},\ and\ \citenamefont
  {Pouquet}}]{rorai2014turbulence}%
  \BibitemOpen
  \bibfield  {author} {\bibinfo {author} {\bibfnamefont {C.}~\bibnamefont
  {Rorai}}, \bibinfo {author} {\bibfnamefont {P.~D.}\ \bibnamefont {Mininni}},
  \ and\ \bibinfo {author} {\bibfnamefont {A.}~\bibnamefont {Pouquet}},\
  }\href@noop {} {\bibfield  {journal} {\bibinfo  {journal} {Phys.\ Rev.\ E}\
  }\textbf {\bibinfo {volume} {89}},\ \bibinfo {pages} {043002} (\bibinfo
  {year} {2014})}\BibitemShut {NoStop}%
\bibitem [{\citenamefont {Khapko}\ \emph {et~al.}(2016)\citenamefont {Khapko},
  \citenamefont {Schlatter}, \citenamefont {Duguet},\ and\ \citenamefont
  {Henningson}}]{khapko2016turbulence}%
  \BibitemOpen
  \bibfield  {author} {\bibinfo {author} {\bibfnamefont {T.}~\bibnamefont
  {Khapko}}, \bibinfo {author} {\bibfnamefont {P.}~\bibnamefont {Schlatter}},
  \bibinfo {author} {\bibfnamefont {Y.}~\bibnamefont {Duguet}}, \ and\ \bibinfo
  {author} {\bibfnamefont {D.~S.}\ \bibnamefont {Henningson}},\ }\href@noop {}
  {\bibfield  {journal} {\bibinfo  {journal} {J.\ Fluid Mech.}\ }\textbf
  {\bibinfo {volume} {795}},\ \bibinfo {pages} {356} (\bibinfo {year}
  {2016})}\BibitemShut {NoStop}%
\bibitem [{\citenamefont {Duguet}\ \emph {et~al.}(2011)\citenamefont {Duguet},
  \citenamefont {Le~Maitre},\ and\ \citenamefont
  {Schlatter}}]{duguet2011stochastic}%
  \BibitemOpen
  \bibfield  {author} {\bibinfo {author} {\bibfnamefont {Y.}~\bibnamefont
  {Duguet}}, \bibinfo {author} {\bibfnamefont {O.}~\bibnamefont {Le~Maitre}}, \
  and\ \bibinfo {author} {\bibfnamefont {P.}~\bibnamefont {Schlatter}},\
  }\href@noop {} {\bibfield  {journal} {\bibinfo  {journal} {Phys.\ Rev.\ E}\
  }\textbf {\bibinfo {volume} {84}},\ \bibinfo {pages} {066315} (\bibinfo
  {year} {2011})}\BibitemShut {NoStop}%
\bibitem [{\citenamefont {Pershin}\ \emph {et~al.}(2019)\citenamefont
  {Pershin}, \citenamefont {Beaume},\ and\ \citenamefont
  {Tobias}}]{pershin2019dynamics}%
  \BibitemOpen
  \bibfield  {author} {\bibinfo {author} {\bibfnamefont {A.}~\bibnamefont
  {Pershin}}, \bibinfo {author} {\bibfnamefont {C.}~\bibnamefont {Beaume}}, \
  and\ \bibinfo {author} {\bibfnamefont {S.~M.}\ \bibnamefont {Tobias}},\
  }\href@noop {} {\bibfield  {journal} {\bibinfo  {journal} {J.\ Fluid Mech.}\
  }\textbf {\bibinfo {volume} {867}},\ \bibinfo {pages} {414} (\bibinfo {year}
  {2019})}\BibitemShut {NoStop}%
\bibitem [{\citenamefont {Canuto}\ \emph {et~al.}(2007)\citenamefont {Canuto},
  \citenamefont {Hussaini}, \citenamefont {Quarteroni},\ and\ \citenamefont
  {Zang}}]{canuto2007spectral}%
  \BibitemOpen
  \bibfield  {author} {\bibinfo {author} {\bibfnamefont {C.}~\bibnamefont
  {Canuto}}, \bibinfo {author} {\bibfnamefont {M.~Y.}\ \bibnamefont
  {Hussaini}}, \bibinfo {author} {\bibfnamefont {A.}~\bibnamefont
  {Quarteroni}}, \ and\ \bibinfo {author} {\bibfnamefont {T.~A.}\ \bibnamefont
  {Zang}},\ }\href@noop {} {\emph {\bibinfo {title} {Spectral methods:
  fundamentals in single domains}}}\ (\bibinfo  {publisher} {Springer-Verlag,
  Berlin Heidelberg},\ \bibinfo {year} {2007})\BibitemShut {NoStop}%
\bibitem [{\citenamefont {Jim{\'e}nez}\ and\ \citenamefont
  {Pinelli}(1999)}]{jimenez1999autonomous}%
  \BibitemOpen
  \bibfield  {author} {\bibinfo {author} {\bibfnamefont {J.}~\bibnamefont
  {Jim{\'e}nez}}\ and\ \bibinfo {author} {\bibfnamefont {A.}~\bibnamefont
  {Pinelli}},\ }\href@noop {} {\bibfield  {journal} {\bibinfo  {journal} {J.\
  Fluid Mech.}\ }\textbf {\bibinfo {volume} {389}},\ \bibinfo {pages} {335}
  (\bibinfo {year} {1999})}\BibitemShut {NoStop}%
\bibitem [{\citenamefont {Jimenez}(2022)}]{jimenez2022streak}%
  \BibitemOpen
  \bibfield  {author} {\bibinfo {author} {\bibfnamefont {J.}~\bibnamefont
  {Jimenez}},\ }\href@noop {} {\bibfield  {journal} {\bibinfo  {journal}
  {arXiv:2202.09814}\ } (\bibinfo {year} {2022})}\BibitemShut {NoStop}%
\bibitem [{\citenamefont {Lundbladh}\ and\ \citenamefont
  {Johansson}(1991)}]{lundbladh1991direct}%
  \BibitemOpen
  \bibfield  {author} {\bibinfo {author} {\bibfnamefont {A.}~\bibnamefont
  {Lundbladh}}\ and\ \bibinfo {author} {\bibfnamefont {A.~V.}\ \bibnamefont
  {Johansson}},\ }\href@noop {} {\bibfield  {journal} {\bibinfo  {journal} {J.\
  Fluid Mech.}\ }\textbf {\bibinfo {volume} {229}},\ \bibinfo {pages} {499}
  (\bibinfo {year} {1991})}\BibitemShut {NoStop}%
\bibitem [{\citenamefont {Van~Saarloos}(2003)}]{van2003front}%
  \BibitemOpen
  \bibfield  {author} {\bibinfo {author} {\bibfnamefont {W.}~\bibnamefont
  {Van~Saarloos}},\ }\href@noop {} {\bibfield  {journal} {\bibinfo  {journal}
  {Physics Reports}\ }\textbf {\bibinfo {volume} {386}},\ \bibinfo {pages} {29}
  (\bibinfo {year} {2003})}\BibitemShut {NoStop}%
\bibitem [{\citenamefont {Hinrichsen}(2000)}]{hinrichsen2000non}%
  \BibitemOpen
  \bibfield  {author} {\bibinfo {author} {\bibfnamefont {H.}~\bibnamefont
  {Hinrichsen}},\ }\href@noop {} {\bibfield  {journal} {\bibinfo  {journal}
  {Advances in {Physics}}\ }\textbf {\bibinfo {volume} {49}},\ \bibinfo {pages}
  {815} (\bibinfo {year} {2000})}\BibitemShut {NoStop}%
\bibitem [{\citenamefont {L{\"u}beck}(2004)}]{lubeck2004universal}%
  \BibitemOpen
  \bibfield  {author} {\bibinfo {author} {\bibfnamefont {S.}~\bibnamefont
  {L{\"u}beck}},\ }\href@noop {} {\bibfield  {journal} {\bibinfo  {journal}
  {Int.\ J.\ Mod.\ Phys.\ B}\ }\textbf {\bibinfo {volume} {18}},\ \bibinfo
  {pages} {3977} (\bibinfo {year} {2004})}\BibitemShut {NoStop}%
\end{thebibliography}%
\end{document}